\documentclass[
    reprint,
    superscriptaddress,
    amsmath, amssymb,
    aps,
    prl,
    floatfix,
]{revtex4-2}

\usepackage{newtxtext, newtxmath}
\usepackage[T1]{fontenc}
\usepackage[usenames, dvipsnames]{xcolor}
\usepackage{graphicx}
\usepackage{dcolumn} 
\usepackage{xpatch}
    \makeatletter \xpretocmd \start@align{\linenomathWithnumbers}{}{\fail}
\usepackage{mathtools, url}
\usepackage{multirow}
\usepackage{CJKutf8}
\usepackage[toc]{appendix}
\usepackage[ruled, linesnumbered]{algorithm2e}
\usepackage[unicode=true, bookmarks=true, bookmarksnumbered=false, bookmarksopen=false, breaklinks=false, pdfborder={0 0 1}, backref=false, colorlinks=false, hidelinks]{hyperref}
    \hypersetup{linkcolor=black, urlcolor=black, citecolor=black, pdfstartview={FitH}}

\usepackage{xr}
\usepackage{calc}

\usepackage[math]{cellspace}
\DeclareSymbolFont{CMlargesymbols}{OMX}{cmex}{m}{n}
\let\sumop\relax\let\prodop\relax
\DeclareMathSymbol{\sumop}{\mathop}{CMlargesymbols}{"50}
\DeclareMathSymbol{\prodop}{\mathop}{CMlargesymbols}{"51}

\SetSymbolFont{operators}{normal}{OT1}{ntxtlf}{m}{n}
\SetSymbolFont{operators}{bold}{OT1}{ntxtlf}{b}{n}

\renewcommand{\vec}[1]{\boldsymbol{#1}}

\newcommand{\p}{\partial}
\newcommand{\dif}{\mathop{}\!\mathrm{d}}
\newcommand{\bn}{\vec{\nabla}}

\newcommand{\ee}{\mathrm{e}}
\newcommand{\ii}{\mathrm{i}}

\renewcommand{\ge}{\geqslant}

\newcommand{\ket}[1]{| #1 \rangle}
\newcommand{\bra}[1]{\langle #1 |}

\allowdisplaybreaks

\begin{document}

\title{Quantum many-body framework for passive-scalar turbulence}

\author{Zhaoyuan Meng}
    \affiliation{State Key Laboratory of Nonlinear Mechanics, Institute of Mechanics, Chinese Academy of Sciences, Beijing 100190, China}
\author{Long Wang}
    \affiliation{State Key Laboratory of Nonlinear Mechanics, Institute of Mechanics, Chinese Academy of Sciences, Beijing 100190, China}
    \affiliation{School of Engineering Sciences, University of Chinese Academy of Sciences, Beijing 100049, China}
\author{Guowei He}
    \email{hgw@lnm.imech.ac.cn}
    \affiliation{State Key Laboratory of Nonlinear Mechanics, Institute of Mechanics, Chinese Academy of Sciences, Beijing 100190, China}
    \affiliation{School of Engineering Sciences, University of Chinese Academy of Sciences, Beijing 100049, China}

\date{\today}

\begin{abstract}
How spatial structures manifest in multi-time correlations is a fundamental question in turbulence.
We develop a non-Hermitian bosonic framework that unifies equal-time anomalous statistics and their temporal propagation within a common operator representation.
A continuous Wegner flow reorganizes stochastic mode couplings in an extended wave-frequency space.
Applied to the Taylor-Kraichnan model, the framework recovers the established equal-time hierarchy and constructs multi-time contributions through propagators acting on successively smaller sets of active fields.
This construction separates uniform transport from intrinsic relative dynamics and shows how a higher-order equal-time state evolves under a generator acting only on the fields that remain dynamically active.
Within an isotropic radial closure, we derive an explicit fourth-order two-time zero mode scaling function that describe how relative dispersion progressively weakens sensitivity to the initial separation.
The framework thus connects spatial intermittency to temporal evolution by identifying how the sequence of observation times determines the propagation of equal-time anomalous structures.
\end{abstract}

\maketitle

\begin{figure*}[t!]
    \centering
    \includegraphics{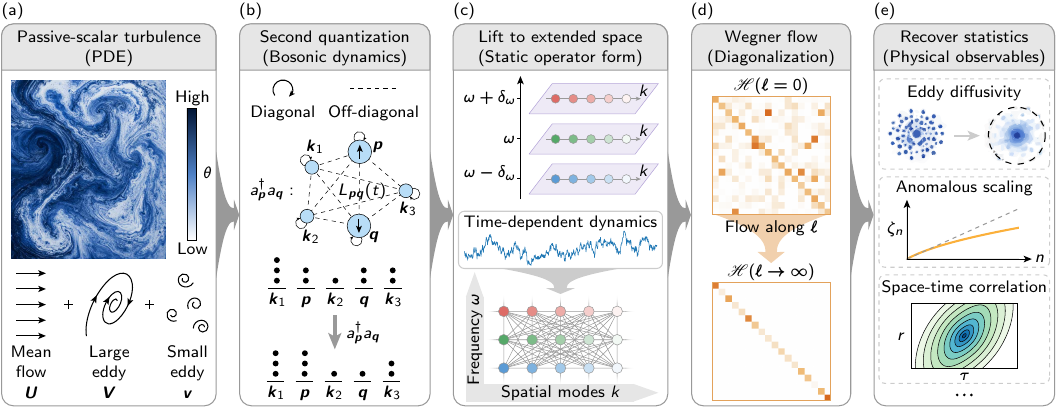}
    \caption{Schematic of the non-Hermitian Wegner flow for passive-scalar turbulence.
    The scalar modes of passive-scalar turbulence, governed by Eq.~\eqref{eq:passive_eq_phys_en} in (a), are mapped to second-quantized bosonic modes in (b).
    The diagonal terms govern the self-evolution of individual modes, while the off-diagonal couplings $L_{\vec{pq}}(t)$ ($\vec{p}\ne\vec{q}$) transfer a bosonic mode from $\vec{q}$ to $\vec{p}$ with a strength determined by $\vec{v}$.
    (c) The time-dependent bosonic dynamics is lifted to an extended space spanned by the spatial mode $\vec{k}$ and frequency $\omega$, where temporal fluctuations manifest as couplings between different frequency sectors.
    The original time-dependent problem is thereby recast as a static algebraic problem in this extended $(\vec{k},\omega)$ space.
    (d) A non-Hermitian Wegner flow continuously suppresses the off-diagonal couplings, driving the extended operator $\mathscr{H}$ toward a diagonal form.
    (e) From this diagonal representation, we systematically reconstruct passive-scalar statistics, e.g., equal-time and multi-time correlations.}
    \label{fig:schematic}
\end{figure*}

\paragraph{Introduction---}
Connecting the spatial organization of fluctuations to their temporal evolution is a central problem in nonequilibrium physics.
In turbulence, transport and deformation jointly shape fluctuations across scales~\cite{Frisch1998_Intermittency, Celani2000_Universality, Gotoh2015_Power, Zhu2023_Circulation, Yao2024_Forward}.
Passive-scalar turbulence offers a canonical setting for studying this connection, with applications to the mixing of temperature, chemical species, and pollutants~\cite{Warhaft2000_Passive, Shraiman2000_Scalar, Dimotakis2005_Turbulent}.
Its intermittent spatial statistics are understood well in the Kraichnan model~\cite{Kraichnan1968_Small, Kraichnan1994_Anomalous, Shraiman1994_Lagrangian, Gawędzki1995_Anomalous, Chertkov1995_Normal, Bernard1996_Anomalous, Fairhall1997_Direct, Falkovich2001_Particles}, where anomalous scaling arises from zero modes of multi-particle transport operators~\cite{Frisch1998_Intermittency, Celani2000_Universality, Gotoh2015_Power}.
These modes encode the geometry of intermittent structures~\cite{Celani2001_Statistical, Arad2001_Statistical}.
Unequal-time measurements additionally probe their advection, sweeping, and deformation~\cite{Kraichnan1964_Kolmogorov, Tennekes1975_Eulerian, Chaves2001_Universal, He2006_Elliptic, Wilczek2012_Wavenumber, He2017_Space, Gorbunova2021_Eulerian}.
Existing dynamics-scaling theories characterize important temporal exponents~\cite{Lvov1997_Temporal, Mitra2004_Varieties, Mitra2005_Dynamics, Ray2008_Universality}, but these exponents alone do not determine the full dependence on observation times and spatial configuration.
The task is therefore to connect equal-time anomalous states to their temporal propagation across different correlation orders and time assignments.

We address this task through the operator connection between classical stochastic dynamics and quantum many-body methods~\cite{Martin1973_Statistical, Andreanov2006_Field}.
Second quantization represents products of scalar fields in particle-number sectors~\cite{Doi1976_Second, Mattis1998_The, delRazo2026_Field}, while continuous flow equation reorganizes stochastic mode couplings into effective interactions~\cite{Wegner1994_Flow, Głazek1993_Renormalization, Verdeny2013_Accurate, Thomson2024_Unravelling}.
Together, these methods place the equal-time boundary state and its subsequent propagation within a common operator framework.
The correlation order determines the particle-number sector of the boundary state, whereas the observation times determine which field factors remain active in each interval.
Averaging their joint propagation retains the correlated displacements induced by the shared velocity field.
These correlations generate pair couplings that preserve statistical information lost when the single-particle propagators are averaged separately.

We demonstrate this framework using the Taylor-Kraichnan (TK) model, which combines spatially rough Gaussian white-in-time advection with a mean flow and frozen random sweeping~\cite{Wang2025_Solvable}.
Recovering the established equal-time hierarchy provides a consistency check.
The unequal-time construction then separates uniform transport from the remaining dynamics and organizes the propagation of anomalous states through generators selected by the observation times.
As the simplest non-Gaussian example, the fourth-order $(2,2)$ sector couples a four-particle equal-time boundary state to an active two-particle generator.
Within an isotropic radial closure, we derive an explicit scaling function showing how relative dispersion progressively weakens the spatial dependence inherited from the equal-time state.
These results identify how temporal measurements probe anomalous spatial structures through the interplay of boundary-state geometry and active-particle dynamics.

\paragraph{Bosonic representation of passive-scalar turbulence---}
We consider the TK model~\cite{Wang2025_Solvable} as a concrete example to illustrate how a partial differential equation (PDE) is lifted to a second-quantized representation in Fig.~\ref{fig:schematic}.
The original stochastic PDE is recast as a non-Hermitian quadratic operator acting on bosonic modes~\cite{Doi1976_Second}.
By extending the Hilbert space to include frequency degrees of freedom, the time evolution is equivalently reformulated as a purely static algebraic constraint~\cite{Howland1974_Stationary}.

We consider the $d$-dimensional passive scalar equation~\cite{Warhaft2000_Passive}
\begin{equation}\label{eq:passive_eq_phys_en}
    \partial_t \theta + \vec u \cdot \bn \theta
    = \varkappa \nabla^2 \theta + f
\end{equation}
on a periodic domain $\mathbb{T}^d$, where $\theta$ is the scalar field, $\varkappa>0$ is the molecular diffusivity, $f$ is an external source, and the velocity field is decomposed as $\vec u(\vec x,t) = \vec U + \vec V + \vec v(\vec x,t)$.
Here, $\vec U$ is a constant vector representing the mean flow, $\vec V$ is a spatially uniform, frozen random sweeping velocity with isotropic Gaussian statistics $\langle V_i \rangle = 0$ and $\langle V_i V_j \rangle = V_0^2 \delta_{ij}$, with $V_0^2$ denoting the variance of each component, and $\vec v(\vec x,t)$ is an isotropic, incompressible, zero-mean Gaussian small-scale velocity with the white-in-time correlation $\langle v_i(\vec x,t) v_j(\vec x',t') \rangle = 2 D_{ij}(\vec x-\vec x') \delta(t-t')$.
The spatial correlation tensor is $D_{ij}(\vec r) = D_0 \delta_{ij} - D_1 [ (d-1+\xi)\delta_{ij} - \xi \frac{r_i r_j}{r^2} ] r^\xi$, where $D_0>0$ denotes the variance at zero separation, $D_1>0$ sets the amplitude of velocity increments, and $\xi \in (0,2)$ is the scaling exponent~\cite{Kraichnan1968_Small}.
The forcing $f$ is a zero-mean, temporally Gaussian white noise with correlation $\langle f(\vec x,t) f(\vec x',t') \rangle = \chi(\vec x-\vec x') \delta(t-t')$, where $\chi(\vec r)$ is smooth, isotropic, and short-ranged over a forcing scale $L_f$.  
Its value at zero separation, $\chi(\vec 0) > 0$, determines the characteristic injection rate of scalar variance.

Applying the spatial Fourier transform $\mathcal{F}$ to Eq.~\eqref{eq:passive_eq_phys_en} yields $\partial_t \hat{\theta}_{\vec k}(t) = \sum_{\vec p} L_{\vec k \vec p}(t) \hat{\theta}_{\vec p}(t) + \hat{f}_{\vec k}(t)$, with the mode-coupling operator $L_{\vec k \vec p}(t) = \delta_{\vec k \vec p} \left[ -\varkappa k^2 - \ii \vec k \cdot (\vec U + \vec V) \right] - \ii \vec k \cdot \hat{\vec v}_{\vec k - \vec p}(t)$.
In this representation, the contributions from molecular diffusion, mean advection, and large-scale sweeping remain diagonal, whereas the small-scale turbulent velocity drives off-diagonal couplings between distinct Fourier modes.

We then map the Fourier-space dynamics to a second-quantized representation.
Since contributions from $f$ can be incorporated via Green's function methods~\cite{Falkovich2001_Particles}, we focus on the unforced propagation.  
We introduce standard bosonic operators $a_{\vec k}$ and $a_{\vec k}^\dagger$ satisfying $[a_{\vec k}, a_{\vec p}^\dagger] = \delta_{\vec k \vec p}$~\cite{Doi1976_Second}.
Lifting the single-particle stochastic generator to Fock space via its particle-number-conserving, normally ordered representation yields the non-Hermitian Hamiltonian
\begin{equation}\label{eq:HTK_en}
    H(t) = \ii \sum_{\vec p, \vec q} L_{\vec p \vec q}(t) a_{\vec p}^\dagger a_{\vec q}.
\end{equation}
Using the bosonic commutation relations, we obtain $\ii [H(t), a_{\vec k}] = \sum_{\vec p} L_{\vec k \vec p}(t) a_{\vec p}$.
Consequently, the Heisenberg equation $\partial_t a_{\vec k} = \ii [H(t), a_{\vec k}]$ recovers the unforced Fourier-space dynamics.

To absorb the time derivative into the operator formulation, we introduce a frequency-extended operator space, as illustrated in Fig.~\ref{fig:schematic}(c).
In this representation, the temporal direction is incorporated into the operator space, and its conjugate variable $\omega$ converts the time-evolution problem into a static constraint in the extended wave-frequency space.
In this space, we define $\mathscr{H} := \ii \partial_t \otimes \mathbb{I} - \ii L(t)$, where $\mathbb{I}$ is the identity operator and $L(t)$ is the single-particle evolution matrix with Fourier-space elements $L_{\vec k\vec p}(t)$.
Accordingly, the Heisenberg equation $\partial_t a_{\vec k} = \ii [H(t), a_{\vec k}]$ is recast as the algebraic constraint
\begin{equation}
    \mathscr{H}\ket{a} = 0,
\end{equation} 
which encodes the source-free passive-scalar dynamics.
Here, the operator-valued vector $\ket{a}=\sum_{\vec{k}} \int_{-\infty}^\infty \frac{\dif\omega}{2\pi} a_{\vec{k},\omega}\ket{\vec{k},\omega}$ is defined in an extended single-particle space spanned by $\ket{\vec k,\omega}:=\ket{\vec k}\otimes\ket{\omega}$. 
These basis states satisfy the orthogonality relation $\langle\vec k,\omega|\vec p,\omega'\rangle =2\pi\delta_{\vec k\vec p}\delta(\omega-\omega')$.
Further details are provided in Supplementary Material (SM)~\cite{SM}.

\paragraph{Non-Hermitian Wegner flow---}
The bosonic representation of Eq.~\eqref{eq:passive_eq_phys_en} yields a quadratic, number-conserving, and non-Hermitian generator.
In the extended wave-frequency space, we decompose $\mathscr H = \mathscr H_{\mathrm{d}} + \mathscr H_{\mathrm{od}}$, with $\mathscr H_{\mathrm{d}}= \sum_{\vec k} \int_{-\infty}^{\infty} \frac{\dif\omega}{2\pi} \lambda_{\vec k,\omega} \ket{\vec k,\omega} \bra{\vec k,\omega}$ and $\mathscr H_{\mathrm{od}} = \sum_{\vec k,\vec p} \int_{-\infty}^{\infty} \frac{\dif\omega}{2\pi} \int_{-\infty}^{\infty} \frac{\dif\omega'}{2\pi} W_{\vec k\omega,\vec p\omega'} \ket{\vec k,\omega} \bra{\vec p,\omega'}$, to separate the propagation of individual modes from the stochastic mode coupling between them.
The diagonal elements $\lambda_{\vec k,\omega} =\omega-\vec k\cdot(\vec U+\vec V)+\ii\varkappa k^2$ describe uniform advection and molecular diffusion.
The off-diagonal elements $W_{\vec k\omega,\vec p\omega'} =-\vec k\cdot\hat{\vec v}_{\vec k-\vec p,\omega-\omega'}$ describe scattering by fluctuating velocity, which transfers scalar amplitude between wave-vector and frequency sectors.

We introduce a continuous parameter $\ell\ge 0$ and suppress off-diagonal couplings using a non-Hermitian Wegner flow~\cite{Wegner1994_Flow, Głazek1993_Renormalization}
\begin{equation}
    \begin{dcases}
        \frac{\dif\mathscr H(\ell)}{\dif\ell} = [\eta(\ell),\mathscr H(\ell)], \\
        \mathscr{H}(0) = \mathscr{H},
    \end{dcases}
	\label{eq:main_wegner_flow}
\end{equation}
with $\eta(\ell) = [\mathscr H_{\mathrm d}^{\dagger}(\ell), \mathscr H_{\mathrm{od}}(\ell)]$.
For composite indices $\alpha=(\vec k,\omega)$ and $\beta=(\vec p,\omega')$, the leading-order flow is $\dif W_{\alpha\beta}/\dif\ell = -|\lambda_\alpha-\lambda_\beta|^2 W_{\alpha\beta}$.
Fixing the diagonal elements at their zeroth-order values yields $W_{\alpha\beta}(\ell) = W_{\alpha\beta}(0) \ee^{ -|\lambda_\alpha-\lambda_\beta|^2\ell }$, showing exponential suppression of nonresonant couplings along the Wegner flow, as illustrated in Fig.~\ref{fig:schematic}(d).
The resulting diagonal normal form is
\begin{equation}
	\varLambda_\alpha
	=
	\lambda_\alpha
	+
	\sum_{\gamma\ne\alpha}
	\frac{W_{\alpha\gamma}W_{\gamma\alpha}}{\lambda_\alpha-\lambda_\gamma},
	\label{eq:main_single_particle_normal_form}
\end{equation}
which defines the effective single-particle generator $\mathscr{H}_{\mathrm{eff}} = \sum_\alpha \varLambda_\alpha \ket{\alpha}\bra{\alpha}$.

The normal form in Eq.~\eqref{eq:main_single_particle_normal_form} determines the scalar propagator, which connects field evolution to correlation functions through ensemble averaging.
Propagating one factor of a given equal-time covariance yields the second-order, two-time correlation.
For several evolving factors, we extend the construction to their joint propagation before averaging over the shared velocity field $\vec{v}$.
Correlated velocity fluctuations then generate deterministic pair couplings that are absent from independently averaged single-particle propagators (see derivation in SM~\cite{SM}).
These couplings describe correlated particle displacements, linking stochastic mode coupling to the relative dispersion governing higher-order scalar statistics.

\paragraph{Equal-time hierarchy---}
The equal-time sector both verifies the many-body normal form and supplies the boundary state for unequal-time evolution.
We show that the Kraichnan hierarchy emerges from the effective many-body operator.
Here we consider the equal-time correlation function $C_n^{\mathrm{eq}}(\vec x_1,\cdots,\vec x_n;t) :=\langle \prod_{j=1}^n \theta(\vec x_j,t)\rangle$.
As detailed in SM~\cite{SM}, we obtain the effective $n$-particle operator $\mathscr H_{\mathrm{eff}}^{(n)} = \ii\partial_t + \ii(\vec U+\vec V)\cdot \sum_{r=1}^{n}\bn_{\vec x_r} - \ii\varkappa \sum_{r=1}^{n} \nabla_{\vec x_r}^{2} - \ii\sum_{r,s=1}^{n} D_{ij}(\vec x_r-\vec x_s) \partial_{x_r^i}\partial_{x_s^j}$ and the corresponding hierarchy equation $\mathscr H_{\mathrm{eff}}^{(n)} C_n^{\mathrm{eq}} = \ii\varPhi_n$, where $\varPhi_n$ accounts for pair contractions of the scalar forcing.

Statistical homogeneity implies invariance under a translation of all coordinates, yielding $\sum_{r=1}^{n} \bn_{\vec x_r}C_n^{\mathrm{eq}} = 0$.
Consequently, the advection term and the zero-separation covariance act solely on the center-of-mass coordinate and vanish.
With $d_{ij}(\vec r) = D_{ij}(\vec 0)-D_{ij}(\vec r)$, the relative generator becomes $\mathcal M_n = \varkappa\sum_{r=1}^{n}\nabla_{\vec x_r}^{2} - \sum_{r,s=1}^{n} d_{ij}(\vec x_r-\vec x_s) \partial_{x_r^i}\partial_{x_s^j}$.
In the stationary inertial range, where molecular diffusion is negligible, the anomalous contributions are governed by homogeneous zero modes satisfying
\begin{equation}
	\mathcal M_n Z_n
	=
	0,
	\label{eq:main_equal_time_zero_mode}
\end{equation}
with $Z_n(\lambda\vec x_1,\cdots,\lambda\vec x_n)=\lambda^{\zeta_n} Z_n(\vec x_1,\cdots,\vec x_n)$ for any dimensionless dilation factor $\lambda>0$.
The homogeneity degree $\zeta_n$ determines the anomalous equal-time scaling exponent of the $n$th-order scalar structure function.
The construction thus recovers the Kraichnan hierarchy as the many-body normal form of the stochastic mode coupling~\cite{Kraichnan1968_Small, Falkovich2001_Particles}.
In this representation, the zero modes encode statistical invariants of relative particle motion and appear as null states of the effective many-particle generator.

\paragraph{Unequal-time normal form---}
The unequal-time sector constitutes the dynamical extension of the equal-time many-body hierarchy.
Here, we show that distinct observation times partition the evolution into ordered intervals, within each of which only the particles associated with future observations remain dynamically active.
This causal organization yields a time-ordered normal form that decouples mean advection, random sweeping, and center-of-mass dispersion from the intrinsic evolution of relative particle configurations.
This decoupling is physically significant because it identifies equal-time zero modes as boundary states for unequal-time dynamics, with lower-particle active generators dictating their subsequent temporal evolution.
Unequal-time anomalous statistics therefore emerge from the propagation of established spatial structures rather than from independent phenomenological assumptions regarding temporal scaling.

Consider $M$ ordered times $t_1<t_2<\cdots<t_M$.
The $m$th time slice contains $n_m$ scalar fields with spatial coordinates $X_m=(\vec x_{m,1},\cdots,\vec x_{m,n_m})$, yielding a total correlation order of $n=\sum_{m=1}^{M}n_m$.
We then define the multi-time Eulerian correlation $C_{\boldsymbol n}^{\mathrm E}(X_M,t_M;\cdots;X_1,t_1) = \langle \prod_{m=1}^{M} \prod_{r=1}^{n_m} \theta(\vec x_{m,r},t_m) \rangle$, where $\boldsymbol n=(n_M,\cdots,n_1)$ specifies the number of fields on each time slice.
For homogeneous correlations, the Fourier amplitudes are restricted to the total-momentum sector $\sum_{m=1}^{M}\sum_{r=1}^{n_m}\vec k_{m,r}=\vec 0$.

The operator structure changes qualitatively when scalar factors occupy distinct observation times.
For $m=2,\cdots,M$, we define the active set $A_m=\bigcup_{j=m}^{M}X_j$, which contains all scalar factors observed at or after $t_m$, together with the time increment $\Delta t_m=t_m-t_{m-1}$.
The source-free multi-time Eulerian correlation is propagated via the time-ordered sequence
\begin{equation}
	C_{\boldsymbol n}^{\mathrm{E}}
	=
	\prod_{m=2}^{M}
	\exp\left(
	\Delta t_m
	\mathcal L_{A_m}
	\right)
	C_n^{\mathrm{eq}}(X_M,\cdots,X_1;t_1),
	\label{eq:main_active_set_propagation}
\end{equation}
where the product is ordered from right to left, and $\mathcal L_A$ is the corresponding Eulerian real-time generator (detailed in SM~\cite{SM}).

For the $m$th interval, we define the total wave vector $\vec K_m=\sum_{i=m}^{M}\sum_{r=1}^{n_i}\vec k_{i,r}$ and its time-weighted counterpart $\vec Q=\sum_{m=1}^{M}\sum_{r=1}^{n_m}\vec k_{m,r}t_m$.
The Fourier coefficients of $C_{\boldsymbol n}^{\mathrm{E}}$ then factorize as
\begin{equation}
	\hat{C}_{\boldsymbol n}^{\mathrm{E}}
	=
    \ee^{-\ii\vec{U}\cdot\vec{Q} - V_0^2|\vec{Q}|^2/2}
	\prod_{m=2}^{M} \ee^{-D_0 K_m^2 \Delta t_m}
	\hat{C}_{\boldsymbol n}^{\mathrm{E, rel}},
	\label{eq:main_multitime_factorization}
\end{equation}
where $C_{\boldsymbol n}^{\mathrm{E, rel}} := \prod_{m=2}^M \ee^{\mathcal{M}_{A_m} \Delta t_m} C_n^{\mathrm{eq}}$ denotes the relative part.
The first factor in Eq.~\eqref{eq:main_multitime_factorization} accounts for deterministic advection and frozen random sweeping.
The product describes the center-of-mass diffusion generated by $D_{ij}(\vec0)=D_0\delta_{ij}$.

For two observation times, we define the future and past coordinates as $X=(\boldsymbol{x}_1,\cdots,\boldsymbol{x}_p)$ and $Y=(\boldsymbol{y}_1,\cdots,\boldsymbol{y}_q)$, respectively, such that $A_2=X$.
The zero-mode contribution in Eq.~\eqref{eq:main_multitime_factorization} then reduces to
\begin{equation}
	\hat{C}_{p,q}^{\mathrm{E}, \mathrm{zm}}(\tau)
	=
	\ee^{-\ii\vec U\cdot\vec{K}_2\tau - (\frac12 V_0^2 \tau + D_0)K_2^2 \tau}
	\mathcal{F}[
	\ee^{\tau\mathcal M_X}
	Z_{p+q}(X,Y)
    ],
	\label{eq:main_two_time_zero_mode}
\end{equation}
where $\mathcal{M}_X = \mathcal{M}_{A_2}$ and $\vec{K}_2=\sum_{r=1}^p \vec{k}_{2,r} = -\sum_{r=1}^q \vec{k}_{1,r}$.
Equation~\eqref{eq:main_two_time_zero_mode} explicitly separates the full Eulerian zero-mode contribution into collective and intrinsic dynamics.
The unequal-time zero mode inherits its spatial exponent from $Z_{p+q}$, while its temporal dynamics is governed by $\ee^{\tau\mathcal M_X}$.

\begin{figure}[t!]
    \centering
    \includegraphics{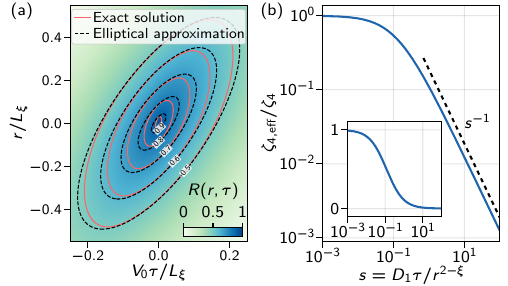}
    \caption{Space-time correlations and dynamic zero-mode scaling.
    (a) Normalized second-order space-time correlation $R(r,\tau)=C_{1,1}^{\mathrm{E}}(r,\tau)/C_2^{\mathrm{eq}}(0)$, showing contours from the exact solution (Eq.~(S159) in SM~\cite{SM}) and the elliptic approximation~\cite{He2006_Elliptic}.
    (b) Effective exponent of the quasi-Lagrangian fourth-order relative zero mode, illustrating the crossover from equal-time anomalous scaling to the $s^{-1}$ regime.
    The parameters are $d=3$, $\xi=4/3$, $\zeta_4=1.05$, $U/V_0=1.5$, and $(\varkappa+D_0)/(V_0L_\xi)=0.02$.}
    \label{fig:fig2}
\end{figure}

For $p=q=1$, from Eq.~\eqref{eq:main_active_set_propagation} we obtain $C_{1,1}(\vec k,\tau) = \ee^{-\ii\vec U\cdot\vec k\tau} \ee^{-\frac12V_0^2k^2\tau^2 - (\varkappa + D_0)k^2\tau} C_2^{\mathrm{eq}}(\vec k)$, with distinct factors describing mean advection, Gaussian sweeping, and single-particle diffusion.
The corresponding space-time correlation in Fig.~\ref{fig:fig2}(a) recovers the analytical results~\cite{Wang2025_Solvable} and closely follows the elliptic approximation~\cite{He2006_Elliptic}.
This agreement establishes the one-particle sector as a benchmark for collective transport before addressing higher-order relative dynamics.

\paragraph{Minimal dynamic zero mode---}
Within the quasi-Lagrangian relative sector, the fourth-order, two-time correlation provides the minimal setting in which an anomalous equal-time zero mode undergoes nontrivial relative evolution.
The second-order Eulerian correlation involves only single-particle propagation and is dominated by convection, sweeping, and diffusion.
By contrast, the $(2,2)$ sector couples a four-particle boundary state to an active two-particle generator.
After removing the center-of-mass motion of the active pair, its evolution directly probes the relative dispersion responsible for the deformation of intermittent scalar structures.

Rather than the full Eulerian scalar-increment correlation, we consider its quasi-Lagrangian relative zero-mode contribution $F_{2,2}^{\mathrm{rel},\mathrm{zm}}(r,\tau) = \mathcal P_r [\ee^{\tau\mathcal M_X}Z_4]$. 
Here, $\mathcal P_r$ is the signed increment projection defined in SM~\cite{SM}, and $\mathcal{M}_X$ acts only on the relative configuration of the two future coordinates after their center-of-mass component is removed.
The full Eulerian correlation additionally contains the collective transport factors in Eq.~\eqref{eq:main_two_time_zero_mode}.
The equal-time four-particle zero mode supplies the boundary state, whereas the active two-particle generator determines its subsequent quasi-Lagrangian evolution.

With collective transport and center-of-mass motion removed, inertial-range scale covariance reduces the quasi-Lagrangian relative dynamics to $F_{2,2}^{\mathrm{rel},\mathrm{zm}}(r,\tau) = A_4r^{\zeta_4} \varPhi_{2,2}(s)$, demonstrating that the unequal-time sector introduces no independent anomalous exponent.
The spatial scaling is inherited from $Z_4$, while the temporal dependence is fixed by the active two-particle relative dynamics.
Here, $s=\tau/\tau_r$ measures the time lag relative to the intrinsic dispersion time $\tau_r=r^z/D_1$, and $z=2-\xi$ denotes the intrinsic relative-dispersion exponent.

Assuming that the increment-projected isotropic sector closes under $\mathcal{M}_X$, the active two-particle dynamics determines an explicit relative zero-mode scaling function.
For $d>z$, with the equal-time condition $\varPhi_{2,2}(0)=1$, we obtain
\begin{equation}
	\varPhi_{2,2}(s)
	=
	\frac{
	\Gamma(
	\frac{d+\zeta_4}{z}
	)
	}{
	\Gamma(
	\frac{d}{z}
	)
	}
	x(s)^{-\zeta_4/z}
	{}_1F_1
	\left(
	-\frac{\zeta_4}{z};
	\frac{d}{z};
	-x(s)
	\right),
	\label{eq:main_Phi_exact}
\end{equation}
with $x(s)=\frac{1}{2(d-1)z^2s}$, the gamma function $\Gamma(\cdot)$, and the confluent hypergeometric function ${}_1F_1(\cdot)$ \cite{Eyink2000_Self} (see SM~\cite{SM} for derivation).
As shown in Fig.~\ref{fig:fig2}(b), the quasi-Lagrangian effective exponent $\zeta_{4,\mathrm{eff}}(r,\tau):=\p\ln F_{2,2}^{\mathrm{rel}, \mathrm{zm}}/\p\ln r$ crosses from $\zeta_{4,\mathrm{eff}}/\zeta_4=1$ for $s \ll 1$ to $\zeta_{4,\mathrm{eff}}/\zeta_4\sim[2d(d-1)zs]^{-1}$ for $s \gg 1$.
In the latter regime, the relative-dispersion scale exceeds $r$, and the zero-mode contribution progressively loses sensitivity to the prescribed separation.
The $s^{-1}$ tail therefore quantifies the loss of relative spatial memory and introduces no additional anomalous exponent.

\paragraph{Discussion---}
We develop a unified operator framework for equal-time anomalous statistics and their unequal-time propagation.
Correlation order determines the particle-number sector of the equal-time boundary state, while the assignment of fields to observation times select the sequence of active generators.
This organization connects different temporal partitions to the same equal-time state and retains the pair couplings induced by a shared velocity field.
The established Kraichnan hierarchy is recovered as the equal-time limit.

The central physical implication is that temporal correlations probe spatial information beyond the equal-time scaling exponent.
This exponent specifies the homogeneity of an anomalous state, whereas its temporal propagation also depends on the geometry of particle configurations.
Different temporal partitions sample that geometry through different propagators and therefore need not yield the same temporal scaling function.
In the $(2,2)$ sector, a two-particle generator propagates a four-particle anomalous boundary state.
Within the isotropic radial closure, we obtain an explicit scaling function that describes how relative dispersion progressively weakens sensitivity to the initial separation.
Its $s^{-1}$ asymptote characterizes the effective spatial exponent rather than the decay of the full correlation.

The framework also provides a starting point for more general transport problems.
Extending it to colored noise requires memory-dependent interactions between propagation intervals intervals~\cite{Zwanzig1961_Memory, Wang2026_Space}.
For Navier-Stokes turbulence, the quadratic convective nonlinearity introduces triadic mode coupling~\cite{Waleffe1992_The, Cambon1999_Linear}.
Combining its second-quantized representation with continuous-flow methods and quantum scrambling~\cite{Thomson2024_Unravelling} offers a direction for developing closures that preserve the symmetries of the underlying dynamics.
The bosonic representation also motivates exploring analog quantum simulation~\cite{Gross2017_Quantum} of turbulent transport, although implementing its non-Hermitian dynamics remains a separate challenge~\cite{Meng2023_Quantum, Meng2024_Simulating, Tennie2025_Quantum}.

\vspace{1em}
\paragraph{Acknowledgments.} 
This work was supported by the National Natural Science Foundation of China (NSFC) through the Excellence Research Group Program for Multiscale Problems in Nonlinear Mechanics (Grant No.~12588201) and the NSFC (Grant No.~12602344).

\paragraph{Data availability.}
No data were created or analyzed in this study.

\bibliographystyle{apsrev4-2.bst}
\bibliography{main.bib}
\end{document}



\title{Supplementary Material for \\ ``Quantum many-body framework for passive-scalar turbulence''}

\author{Zhaoyuan Meng}
    \affiliation{State Key Laboratory of Nonlinear Mechanics, Institute of Mechanics, Chinese Academy of Sciences, Beijing 100190, China}
\author{Long Wang}
    \affiliation{State Key Laboratory of Nonlinear Mechanics, Institute of Mechanics, Chinese Academy of Sciences, Beijing 100190, China}
    \affiliation{School of Engineering Sciences, University of Chinese Academy of Sciences, Beijing, 100049, China}
\author{Guowei He}
    \email{hgw@lnm.imech.ac.cn}
    \affiliation{State Key Laboratory of Nonlinear Mechanics, Institute of Mechanics, Chinese Academy of Sciences, Beijing 100190, China}
    \affiliation{School of Engineering Sciences, University of Chinese Academy of Sciences, Beijing, 100049, China}
    
\date{\today}

\maketitle

\beginsupplement
\renewcommand{\thepage}{S\arabic{page}}
\renewcommand{\citenumfont}[1]{S#1}
\renewcommand{\bibnumfmt}[1]{[S#1]}

\tableofcontents

\section{From the Taylor-Kraichnan model to a bosonic representation}\label{sec:supp_bosonization}
We present a detailed derivation of the bosonic representation for turbulent transport.
Starting from the Taylor-Kraichnan (TK) passive-scalar equation~\cite{Wang2025_Solvable, Kraichnan1968_Small}, we transform it to spectral space and lift the resulting linear stochastic evolution to a second-quantized bosonic form.

\subsection{Taylor-Kraichnan passive-scalar equation}

Consider a passive scalar $\theta(\vec x,t)$ on a $d$-dimensional periodic domain $\mathbb T^d=[0,2\pi)^d$ with $|\mathbb T^d|=(2\pi)^d$.
The passive scalar is advected by a prescribed incompressible velocity field, diffused with molecular diffusivity $\varkappa$, and driven by an external scalar forcing $f$, according to the advection-diffusion equation~\cite{Majda1999_Simplified, Shraiman2000_Scalar, Falkovich2001_Particles}
\begin{equation}\label{eq:supp_scalar_pde}
    \partial_t \theta(\vec x,t)
    +
    \vec u(\vec x,t)\cdot\bn\theta(\vec x,t)
    =
    \varkappa\nabla^2\theta(\vec x,t)+f(\vec x,t).
\end{equation}

We adopt the Fourier series convention
\begin{equation}\label{eq:supp_fourier_theta}
    \theta(\vec x,t)
    =
    \sum_{\vec k}\hat\theta_{\vec k}(t) \ee^{\ii\vec k\cdot\vec x},
    \quad
    \hat\theta_{\vec k}(t)
    =
    \frac{1}{|\mathbb T^d|}
    \int_{\mathbb T^d}
    \theta(\vec x,t)\ee^{-\ii\vec k\cdot\vec x}\dif\vec x,
\end{equation}
for an arbitrary function $\theta$, where $\vec k\in \mathbb Z^d$.
The incompressibility condition then reads $\vec k\cdot\hat{\vec v}_{\vec k}(t)=0$.
To isolate the spatially uniform sweeping component from $\vec v$, we impose $\hat{\vec v}_{\vec 0}(t)= \vec 0$.

Fourier transformation of Eq.~\eqref{eq:supp_scalar_pde}, combined with the incompressibility condition $\vec p\cdot\hat{\vec v}_{\vec k-\vec p} = \vec k\cdot\hat{\vec v}_{\vec k-\vec p}$ yields the spectral representation of the TK model
\begin{equation}\label{eq:supp_fourier_scalar_full}
    \partial_t\hat\theta_{\vec k}(t)
    =
    -\left[
    \varkappa k^2 + \ii\vec k\cdot(\vec U + \vec V) 
    \right]\hat\theta_{\vec k}(t)
    - \ii\sum_{\vec p}
    \vec k\cdot\hat{\vec v}_{\vec k-\vec p}(t)
    \hat\theta_{\vec p}(t)
    +
    \hat f_{\vec k}(t).
\end{equation}
By introducing the decomposition
\begin{equation}\label{eq:supp_L_decomposition}
    L_{\vec k\vec p}(t)
    =
    L^{(0)}_{\vec k\vec p}
    +
    L^{(1)}_{\vec k\vec p}(t),
\end{equation}
we isolate the diagonal contribution
\begin{equation}\label{eq:supp_L0}
    L^{(0)}_{\vec k\vec p}
    =
    -\delta_{\vec k\vec p}
    \left[
    \varkappa k^2+\ii\vec k\cdot(\vec U + \vec V)
    \right],
\end{equation}
from the fluctuation-induced mode coupling
\begin{equation}\label{eq:supp_L1}
    L^{(1)}_{\vec k\vec p}(t)
    =
    -\ii\vec k\cdot\hat{\vec v}_{\vec k-\vec p}(t).
\end{equation}
Consequently, Eq.~\eqref{eq:supp_fourier_scalar_full} takes the compact form
\begin{equation}\label{eq:supp_theta_L_equation}
    \partial_t\hat\theta_{\vec k}(t)
    =
    \sum_{\vec p}
    L_{\vec k\vec p}(t)\hat\theta_{\vec p}(t)
    +
    \hat f_{\vec k}(t).
\end{equation}
Because $\hat{\vec v}_{\vec 0}=\vec 0$, the small-scale velocity contributes exclusively to the mixing of distinct Fourier modes.
The diagonal operator $L^{(0)}$ encompasses molecular diffusion, deterministic mean advection, and frozen sweeping advection.

Since the forcing is additive and can be incorporated using Green's functions, response functions, or an auxiliary linear source operator, we first focus on the homogeneous propagation problem
\begin{equation}\label{eq:supp_homogeneous_theta}
    \partial_t\hat\theta_{\vec k}(t)
    =
    \sum_{\vec p}
    L_{\vec k\vec p}(t)\hat\theta_{\vec p}(t).
\end{equation}
In vector notation, Eq.~\eqref{eq:supp_homogeneous_theta} reads
\begin{equation}
    \partial_t\hat{\vec \theta}(t)
    =
    L(t)\hat{\vec \theta}(t),
    \label{eq:supp_vector_L_equation}
\end{equation}
with a non-Hermitian operator $L(t)$.
Specifically, diffusion contributes a real, negative diagonal part, whereas advection yields imaginary diagonal elements and off-diagonal mode-coupling terms.

\subsection{Bosonic Fock-space representation}

We introduce bosonic annihilation and creation operators, $a$ and $a^\dagger$, satisfying the canonical commutation relations
\begin{equation}\label{eq:supp_boson_commutation}
    [a_{\vec k},a_{\vec p}^{\dagger}]
    =
    \delta_{\vec k\vec p},
    \quad
    [a_{\vec k},a_{\vec p}]
    =
    [a_{\vec k}^{\dagger},a_{\vec p}^{\dagger}]
    =0.
\end{equation}
The vacuum state is defined by $a_{\vec k}|0\rangle=0$ for all $\vec k$, and the one-particle basis states are defined as
\begin{equation}
    |\vec k\rangle
    =
    a_{\vec k}^{\dagger}|0\rangle .
\end{equation}

Next, we define a quadratic operator
\begin{equation}\label{eq:supp_H_def}
    H(t)
    =
    \ii\sum_{\vec p,\vec q}
    L_{\vec p\vec q}(t)
    a_{\vec p}^{\dagger}a_{\vec q}.
\end{equation}
Because $L(t)$ is non-Hermitian, $H(t)$ is also generally non-Hermitian.
Accordingly, this operator should be interpreted as an auxiliary evolution generator rather than a physical Hermitian Hamiltonian.
We define the corresponding one-particle state as
\begin{equation}\label{eq:supp_theta_state}
    |\theta(t)\rangle
    =
    \sum_{\vec k}
    \hat\theta_{\vec k}(t)|\vec k\rangle
    =
    \sum_{\vec k}
    \hat\theta_{\vec k}(t)a_{\vec k}^{\dagger}|0\rangle .
\end{equation}
By applying the identity $a_{\vec p}^{\dagger}a_{\vec q}|\vec k\rangle = \delta_{\vec q\vec k}|\vec p\rangle$, we obtain
\begin{equation}
    H(t)|\theta(t)\rangle
    =
    \ii\sum_{\vec p,\vec q,\vec k}
    L_{\vec p\vec q}(t)
    \hat\theta_{\vec k}(t)
    a_{\vec p}^{\dagger}a_{\vec q}|\vec k\rangle
     \\
    =
    \ii\sum_{\vec p,\vec q}
    L_{\vec p\vec q}(t)
    \hat\theta_{\vec q}(t)
    |\vec p\rangle.
\end{equation}
Thus, the Schrödinger-like equation
\begin{equation}\label{eq:supp_schrodinger_like}
    \ii\partial_t|\theta(t)\rangle
    =
    H(t)|\theta(t)\rangle
\end{equation}
is equivalent to Eq.~\eqref{eq:supp_homogeneous_theta}.

This equivalence can also be formulated directly at the operator level.
The bosonic commutation relations yield
\begin{equation}\label{eq:supp_commutator_a}
    [a_{\vec p}^{\dagger}a_{\vec q},
    a_{\vec k}]
    =
    a_{\vec p}^{\dagger}
    [a_{\vec q}, a_{\vec k}]
    +
    [a_{\vec p}^{\dagger},a_{\vec k}]
    a_{\vec q}
    = -\delta_{\vec p\vec k}a_{\vec q}.
\end{equation}
From this, it follows that
\begin{equation}\label{eq:supp_commutator_H_a}
    [H(t),a_{\vec k}]
    = \ii\sum_{\vec p,\vec q} L_{\vec p\vec q}(t)
    [a_{\vec p}^{\dagger}a_{\vec q}, a_{\vec k}]
    = -\ii\sum_{\vec q} L_{\vec k\vec q}(t)a_{\vec q}.
\end{equation}
As a result, the Heisenberg equation of motion
\begin{equation}\label{eq:supp_heisenberg_convention}
    \partial_t a_{\vec k}(t)
    =
    \ii[H(t),a_{\vec k}(t)]
\end{equation}
reduces to
\begin{equation}\label{eq:supp_heisenberg_equation}
    \partial_t a_{\vec k}(t)
    =
    \sum_{\vec p}
    L_{\vec k\vec p}(t)a_{\vec p}(t),
\end{equation}
which is algebraically equivalent to Eq.~\eqref{eq:supp_homogeneous_theta}.
Thus, the annihilation operators satisfy the same linear evolution equation as the classical Fourier modes $\hat\theta_{\vec k}(t)$.

The mapping from Eq.~\eqref{eq:supp_homogeneous_theta} to Eq.~\eqref{eq:supp_heisenberg_convention} establishes the formal lift of the classical TK passive-scalar model to a non-Hermitian quadratic bosonic dynamics~\cite{Doi1976_Second}.
The subsequent extended-space construction absorbs the explicit time dependence into an algebraic operator, which can then be diagonalized via a non-Hermitian Wegner flow.

\subsection{Lift to the extended space}\label{sec:supp_extended_space}
We next lift the time-dependent bosonic dynamics of Eq.~\eqref{eq:supp_heisenberg_convention} to an extended space.
Absorbing the explicit time derivative into an enlarged operator recasts the original evolution equation as an algebraic constraint in the joint space of spatial modes and frequencies.

We define the time-extended operator
\begin{equation}\label{eq:supp_K_def}
    \mathcal K
    =
    \mathrm{i}\partial_t+H(t).
\end{equation}
Here, the derivative operator $\mathrm{i}\partial_t$ acts on the explicit time dependence of operator-valued functions.
For any time-dependent operator $A(t)$, the commutator with $\mathrm{i}\partial_t$ is defined through its action on an arbitrary state $\ket{\psi(t)}$, yielding
\begin{equation}
    \begin{aligned}
        [\mathrm{i}\partial_t,A(t)] \ket{\psi(t)}
        &= \ii\p_t[A(t)\ket{\psi(t)}] - A(t)\ii\p_t\ket{\psi(t)}
        \\
        &= \ii [\p_t A(t)] \ket{\psi(t)} + \ii A(t) \p_t \ket{\psi(t)} - A(t)\ii\p_t\ket{\psi(t)}
        \\
        &= \ii [\p_t A(t)] \ket{\psi(t)}.
    \end{aligned}
\end{equation}
Consequently, the commutator with $\mathrm{i}\partial_t$ reduces to
\begin{equation}\label{eq:supp_derivative_commutator}
    [\mathrm{i}\partial_t,A(t)]
    =
    \mathrm{i}\partial_t A(t).
\end{equation}
By employing Eq.~\eqref{eq:supp_commutator_H_a}, we obtain
\begin{equation}\label{eq:supp_K_commutator}
    [\mathcal K,a_{\vec k}]
    = [\mathrm{i}\partial_t,a_{\vec k}] + [H(t),a_{\vec k}]
    = \mathrm{i}\partial_t a_{\vec k} - \mathrm{i}\sum_{\vec p} L_{\vec k\vec p}(t)a_{\vec p}
    = \mathrm{i} \bigg[ \partial_t a_{\vec k} - \sum_{\vec p} L_{\vec k\vec p}(t)a_{\vec p} \bigg].
\end{equation}
Therefore, the Heisenberg equation of motion, Eq.~\eqref{eq:supp_heisenberg_convention}, is equivalent to the algebraic commutator constraint
\begin{equation}\label{eq:supp_K_constraint}
    [\mathcal K,a_{\vec k}]=0.
\end{equation}
This relation provides the first formulation of the lift, in which the original time evolution is recast as a zero-commutator condition with the extended operator $\mathcal K$.

It is useful to distinguish the Fock-space operator $\mathcal K$ from the one-particle matrix induced on the annihilation sector.
Denoting this matrix by
\begin{equation}\label{eq:supp_scriptK_def}
    \mathscr{H} := \mathrm{i}\partial_t \otimes \mathbb I - \mathrm{i}L(t),
\end{equation}
the corresponding constraint reads
\begin{equation}\label{eq:supp_scriptK_time_constraint}
    \sum_{\vec p}
    \left[
    \mathrm{i}\delta_{\vec k\vec p}\partial_t
    -
    \mathrm{i}L_{\vec k\vec p}(t)
    \right]
    a_{\vec p}(t)
    =
    0.
\end{equation}

We next transform the temporal coordinate to its conjugate frequency.
For a stochastic velocity field, the associated frequency is continuous.
We adopt the Fourier conventions
\begin{equation}\label{eq:supp_a_time_fourier}
    a_{\vec k}(t) = \int_{-\infty}^{\infty} \frac{\dif\omega}{2\pi} a_{\vec k,\omega} \mathrm{e}^{-\mathrm{i}\omega t},
    \quad
    a_{\vec k,\omega} = \int_{-\infty}^{\infty} \dif t\, a_{\vec k}(t) \,\mathrm{e}^{\mathrm{i}\omega t},
    \quad
    \hat{\vec v}_{\vec q}(t) = \int_{-\infty}^{\infty} \frac{\dif\omega}{2\pi} \hat{\vec v}_{\vec q,\omega} \,\mathrm{e}^{-\mathrm{i}\omega t},
    \quad
    \hat{\vec v}_{\vec q,\omega} = \int_{-\infty}^{\infty} \dif t\, v_{\vec q}(t) \,\mathrm{e}^{\mathrm{i}\omega t}.
\end{equation}
Substituting Eq.~\eqref{eq:supp_a_time_fourier} into Eq.~\eqref{eq:supp_scriptK_time_constraint} yields the frequency-space constraint
\begin{equation}\label{eq:supp_algebraic_constraint_frequency}
    \sum_{\vec p}
    \int_{-\infty}^{\infty}
    \frac{\dif\omega'}{2\pi}
    \mathscr{H}_{\vec k\omega,\vec p\omega'}
    a_{\vec p,\omega'}
    =
    0,
\end{equation}
where the extended-space matrix element is decomposed as
\begin{equation}\label{eq:supp_scriptK_matrix_decomposition}
    \mathscr{H}_{\vec k\omega,\vec p\omega'}
    =
    2\pi \delta_{\vec k\vec p}
    \delta(\omega-\omega') \lambda_{\vec k,\omega}
    +
    W_{\vec k\omega,\vec p\omega'},
\end{equation}
with the diagonal part
\begin{equation}\label{eq:supp_lambda_def}
    \lambda_{\vec k,\omega}
    =
    \omega
    -
    \vec k\cdot (\vec U + \vec V)
    +
    \mathrm{i}\varkappa k^2,
\end{equation}
and the off-diagonal coupling
\begin{equation}\label{eq:supp_W_def}
    W_{\vec k\omega,\vec p\omega'}
    =
    -\vec k\cdot
    \hat{\vec v}_{\vec k-\vec p,\omega-\omega'}.
\end{equation}
Equations~\eqref{eq:supp_algebraic_constraint_frequency} to \eqref{eq:supp_W_def} provide the desired algebraic representation of the original time-dependent dynamics.

\subsection{Extended-space basis}
We next formulate a compact representation of the algebraic constraint within an extended one-particle space.
Let $\mathbb Z^d$ denote the lattice of Fourier wave vectors on the periodic domain $\mathbb T^d=[0,2\pi)^d$.
The spatial-mode Hilbert space is the sequence space $\mathcal H_{\vec k}=\ell^2(\mathbb Z^d)$ spanned by the orthonormal basis $\{|\vec k\rangle\}_{\vec k\in\mathbb Z^d}$ satisfying
\begin{equation}
    \langle \vec k|\vec p\rangle
    =
    \delta_{\vec k\vec p},
    \quad
    \sum_{\vec k\in\mathbb Z^d}
    |\vec k\rangle\langle \vec k|
    =
    \mathbb I_{\vec k}.
\end{equation}
The frequency space is defined as
\begin{equation}
    \mathcal H_{\omega}
    =
    L^2\left(\mathbb R,\frac{\dif\omega}{2\pi}\right),
\end{equation}
with generalized basis vectors $|\omega\rangle$ normalized according to
\begin{equation}
    \langle \omega|\omega'\rangle
    =
    2\pi\delta(\omega-\omega'),
    \quad
    \int_{-\infty}^{\infty}
    \frac{\dif\omega}{2\pi}
    |\omega\rangle\langle\omega|
    =
    \mathbb I_{\omega}.
\end{equation}
The extended one-particle space is then defined as the tensor product
\begin{equation}
    \mathcal H_{\mathrm{ext}}
    =
    \mathcal H_{\vec k}
    \otimes
    \mathcal H_{\omega}.
\end{equation}
Its generalized basis vectors are given by
\begin{equation}
    |\vec k,\omega\rangle
    =
    |\vec k\rangle\otimes|\omega\rangle,
\end{equation}
with the normalization
\begin{equation}
    \langle \vec k,\omega|\vec p,\omega'\rangle
    =
    2\pi\delta_{\vec k\vec p}\delta(\omega-\omega'),
\end{equation}
and the resolution of identity
\begin{equation}
    \mathbb I_{\mathrm{ext}}
    =
    \sum_{\vec k}
    \int_{-\infty}^{\infty}
    \frac{\dif\omega}{2\pi}
    |\vec k,\omega\rangle
    \langle \vec k,\omega|.
\end{equation}

The frequency components of the annihilation operators define a state vector
\begin{equation}
    |a\rangle
    =
    \sum_{\vec k}
    \int_{-\infty}^{\infty}
    \frac{\dif\omega}{2\pi}
    a_{\vec k,\omega}
    |\vec k,\omega\rangle
\end{equation}
in this extended space.
The algebraic constraint corresponding to Eq.~\eqref{eq:supp_scriptK_time_constraint} then takes the form
\begin{equation}\label{eq:supp_extended_H_constraint}
    \mathscr H |a\rangle=0.
\end{equation}
In the $|\vec k,\omega\rangle$ basis, we decompose $\mathscr H$ as
\begin{equation}
    \mathscr H
    =
    \mathscr{H}_{\mathrm d}+\mathscr H_{\mathrm{od}},
\end{equation}
with the diagonal part
\begin{equation}
    \mathscr{H}_{\mathrm d}
    =
    \sum_{\vec k}
    \int_{-\infty}^{\infty}
    \frac{\dif\omega}{2\pi}
    \lambda_{\vec k,\omega}
    |\vec k,\omega\rangle
    \langle \vec k,\omega|,
    \label{eq:supp_scrH0}
\end{equation}
and the off-diagonal coupling
\begin{equation}
    \mathscr{H}_{\mathrm{od}}
    =
    \sum_{\vec k,\vec p}
    \int_{-\infty}^{\infty}
    \frac{\dif\omega}{2\pi}
    \frac{\dif\omega'}{2\pi}
    W_{\vec k\omega,\vec p\omega'}
    |\vec k,\omega\rangle
    \langle \vec p,\omega'|.
    \label{eq:supp_scrH1}
\end{equation}

Under the normalization adopted above, Eq.~\eqref{eq:supp_extended_H_constraint} is equivalent to Eq.~\eqref{eq:supp_algebraic_constraint_frequency}.
Consequently, $\mathscr H$ acts as a static algebraic operator in the extended $(\vec k,\omega)$ space.
The diagonal part $\mathscr{H}_{\dif}$ accounts for molecular diffusion and uniform advection by $\vec U+\vec V$, whereas $\mathscr{H}_{\mathrm{od}}$ governs the scattering among spatial modes and frequency sectors induced by turbulent velocity fluctuations.
The coupling term $W_{\vec k\omega,\vec p\omega'}$ mediates the transfer of a scalar quantum from $(\vec p,\omega')$ to $(\vec k,\omega)$.
The exchanged turbulent velocity mode carries the wave vector $\vec q = \vec k - \vec p$ and frequency $\varOmega = \omega - \omega'$.
Accordingly, temporal fluctuations of the velocity field manifest in the extended space as couplings between different frequency sectors.

\section{Wegner-flow diagonalization for equal-time correlations}

\subsection{Single-particle non-Hermitian Wegner flow}\label{sec:supp_single_particle_wegner}

We diagonalize the single-particle operator $\mathscr H$ within the extended space by employing a non-Hermitian Wegner flow.
This operator acts on the extended one-particle space $\mathcal H_{\mathrm{ext}}=\mathcal H_{\vec k}\otimes\mathcal H_{\omega}$ under the algebraic constraint of Eq.~\eqref{eq:supp_extended_H_constraint}.

By introducing the composite indices $\alpha=(\vec k,\omega)$ and $\beta=(\vec p,\omega')$, we recast Eq.~\eqref{eq:supp_algebraic_constraint_frequency} into the compact form
\begin{equation}\label{eq:supp_composite_constraint}
    \sum_{\beta}
    \mathscr{H}_{\alpha\beta}a_{\beta}=0,
    \quad
    \mathscr{H}_{\alpha\beta}
    =
    \lambda_{\alpha}\delta_{\alpha\beta}
    +
    W_{\alpha\beta},
\end{equation}
where
\begin{equation}\label{S-eq:lambda_alpha}
    \lambda_{\alpha}
    =
    \lambda_{\vec k,\omega}
    =
    \omega
    -
    \vec k\cdot (\vec U + \vec V)
    +
    \mathrm{i}\varkappa k^2
\end{equation}
and
\begin{equation}\label{S-eq:W_alphabeta}
    W_{\alpha\beta}
    =
    W_{\vec k\omega,\vec p\omega'}
    =
    -\vec k\cdot
    \hat{\vec v}_{\vec k-\vec p,\omega-\omega'}.
\end{equation}
In this representation, the summation over the composite index is defined by
\begin{equation}
    \sum_{\alpha}
    :=
    \sum_{\vec k}
    \int_{-\infty}^{\infty}
    \frac{\dif\omega}{2\pi},
\end{equation}
while the corresponding extended-space delta function is given by
\begin{equation}
    \delta_{\alpha\beta}
    :=
    2\pi
    \delta_{\vec k\vec p}
    \delta(\omega-\omega').
\end{equation}
This mapping converts the explicitly time-dependent transport problem into a static, infinite-dimensional algebraic problem within the extended $(\vec k,\omega)$ space.
To diagonalize the algebraic operator $\mathscr H$ via the Wegner flow, we decompose it into diagonal and off-diagonal parts, $\mathscr H = \mathscr H_{\mathrm d} + \mathscr H_{\mathrm{od}}$, where $(\mathscr H_{\mathrm d})_{\alpha\beta} = \lambda_{\alpha}\delta_{\alpha\beta}$ and $(\mathscr H_{\mathrm{od}})_{\alpha\beta} = W_{\alpha\beta}$.
The non-Hermitian flow is designed to continuously suppress the off-diagonal component $\mathscr H_{\mathrm{od}}$, thereby driving $\mathscr H$ into a diagonal, normal form.

We introduce a continuous flow parameter $\ell\geq 0$, expressing the Wegner flow equation as
\begin{equation}\label{eq:supp_wegner_flow_def}
    \begin{dcases}
        \frac{\dif \mathscr H(\ell)}{\dif \ell} = [\eta(\ell),\mathscr H(\ell)], \\
        \mathscr H(0)=\mathscr H .
    \end{dcases}
\end{equation}
For the non-Hermitian operator under consideration, we adopt the generator
\begin{equation}
    \eta(\ell)
    =
    [
    \mathscr H_{\mathrm d}^{\dagger}(\ell),
    \mathscr H_{\mathrm{od}}(\ell)
    ].
    \label{eq:supp_nonhermitian_eta_def}
\end{equation}
While this choice reduces to the standard Wegner generator in the Hermitian limit~\cite{Wegner1994_Flow, Głazek1993_Renormalization}, for a non-Hermitian operator it generally induces a non-unitary similarity transformation.

For any flow parameter $\ell$, the diagonal and off-diagonal components of the operator are $(\mathscr H_{\mathrm d})_{\alpha\beta}(\ell) = \lambda_{\alpha}(\ell)\delta_{\alpha\beta}$ and $(\mathscr H_{\mathrm{od}})_{\alpha\beta}(\ell) = W_{\alpha\beta}(\ell)$, respectively.
Substituting $(\mathscr H_{\mathrm d}^{\dagger})_{\alpha\beta}(\ell) = \lambda_{\alpha}^{*}(\ell)\delta_{\alpha\beta}$ into Eq.~\eqref{eq:supp_nonhermitian_eta_def} yields the off-diagonal matrix elements of the generator
\begin{equation}
    \begin{aligned}
        \eta_{\alpha\beta}(\ell)
        &=
        \sum_{\gamma}
        \left[
        (\mathscr H_{\mathrm d}^{\dagger})_{\alpha\gamma}
        W_{\gamma\beta}
        -
        W_{\alpha\gamma}
        (\mathscr H_{\mathrm d}^{\dagger})_{\gamma\beta}
        \right]
         \\
        &=
        \lambda_{\alpha}^{*}(\ell)W_{\alpha\beta}(\ell)
        -
        W_{\alpha\beta}(\ell)\lambda_{\beta}^{*}(\ell)
         \\
        &=
        \big[
        \lambda_{\alpha}^{*}(\ell)
        -
        \lambda_{\beta}^{*}(\ell)
        \big]
        W_{\alpha\beta}(\ell).
    \end{aligned}
\end{equation}
The diagonal components vanish identically, namely $\eta_{\alpha\alpha}(\ell)=0$.
By defining
\begin{equation}\label{eq:supp_Delta_def}
    \Delta_{\alpha\beta}(\ell)
    =
    \lambda_{\alpha}(\ell)
    -
    \lambda_{\beta}(\ell),
\end{equation}
we express the generator as
\begin{equation}\label{eq:supp_eta_matrix_element}
    \eta_{\alpha\beta}(\ell)
    =
    \Delta_{\alpha\beta}^{*}(\ell)
    W_{\alpha\beta}(\ell),
    \quad
    \alpha\neq\beta.
\end{equation}

Taking the diagonal part of Eq.~\eqref{eq:supp_wegner_flow_def}, we obtain the evolution of the diagonal components as
\begin{equation} 
    \frac{\dif \lambda_{\alpha}}{\dif \ell}
    =
    \left[
    \eta,\mathscr H
    \right]_{\alpha\alpha}.
\end{equation}
Since $\eta_{\alpha\alpha}=0$ and $\mathscr H_{\mathrm d}$ is diagonal, the diagonal elements $[\eta,\mathscr H_{\mathrm d}]_{\alpha\alpha}$ vanish, yielding the flow equation
\begin{equation}
    \frac{\dif \lambda_{\alpha}}{\dif \ell}
    =
    [\eta,\mathscr H_{\mathrm{od}}]_{\alpha\alpha},
\end{equation}
which implies that only the off-diagonal part contributes to the flow.
Expanding this commutator yields
\begin{equation}
    \frac{\dif \lambda_{\alpha}}{\dif \ell}
    =
    \sum_{\gamma}
    \left(
    \eta_{\alpha\gamma}
    W_{\gamma\alpha}
    -
    W_{\alpha\gamma}
    \eta_{\gamma\alpha}
    \right),
\end{equation}
which, upon substituting Eq.~\eqref{eq:supp_eta_matrix_element}, simplifies to
\begin{equation}\label{eq:supp_lambda_flow}
    \frac{\dif \lambda_{\alpha}}{\dif \ell}
    =
    \sum_{\gamma\neq\alpha}
    \left(
    \Delta_{\alpha\gamma}^*
    W_{\alpha\gamma}W_{\gamma\alpha}
    -
    W_{\alpha\gamma}
    \Delta_{\gamma\alpha}^*
    W_{\gamma\alpha}
    \right)
    =
    2
    \sum_{\gamma\neq\alpha}
    \Delta_{\alpha\gamma}^*
    W_{\alpha\gamma}W_{\gamma\alpha}.
\end{equation}
Here, all quantities in Eq.~\eqref{eq:supp_lambda_flow} are evaluated at the same flow parameter $\ell$.

For $\alpha\neq\beta$, the evolution of the off-diagonal components is governed by
\begin{equation}
    \frac{\dif W_{\alpha\beta}}{\dif \ell}
    =
    [\eta,\mathscr H_{\mathrm d}]_{\alpha\beta}
    +
    [\eta,\mathscr H_{\mathrm{od}}]_{\alpha\beta}.
\end{equation}
The contribution from the diagonal component $\mathscr H_{\mathrm d}$ evaluates to
\begin{equation}\label{eq:supp_eta_Hd_ab}
    [\eta,\mathscr H_{\mathrm d}]_{\alpha\beta}
    =
    \sum_{\gamma}
    \left[
    \eta_{\alpha\gamma}
    (\mathscr H_{\mathrm d})_{\gamma\beta}
    -
    (\mathscr H_{\mathrm d})_{\alpha\gamma}
    \eta_{\gamma\beta}
    \right]
    =
    \eta_{\alpha\beta}\lambda_{\beta}
    -
    \lambda_{\alpha}\eta_{\alpha\beta}
    =
    -\Delta_{\alpha\beta}\eta_{\alpha\beta}.
\end{equation}
Substituting Eq.~\eqref{eq:supp_eta_matrix_element} into Eq.~\eqref{eq:supp_eta_Hd_ab} yields
\begin{equation}
    [\eta,\mathscr H_{\mathrm d}]_{\alpha\beta}
    =
    -
    |\Delta_{\alpha\beta}|^2
    W_{\alpha\beta}.
    \label{eq:supp_offdiag_decay_term}
\end{equation}
Meanwhile, the contribution from the off-diagonal component is
\begin{equation}
    [\eta,\mathscr H_{\mathrm{od}}]_{\alpha\beta}
    =
    \sum_{\gamma}
    \left[
    \eta_{\alpha\gamma}W_{\gamma\beta}
    -
    W_{\alpha\gamma}\eta_{\gamma\beta}
    \right].
\end{equation}
Since the diagonal elements vanish ($W_{\alpha\alpha}=0$ and $\eta_{\alpha\alpha}=0$), the terms corresponding to $\gamma=\alpha$ and $\gamma=\beta$ do not contribute, yielding
\begin{equation}\label{eq:supp_offdiag_quadratic_term}
    \begin{aligned}
        [\eta,\mathscr H_{\mathrm{od}}]_{\alpha\beta}
        &=
        \sum_{\gamma\neq\alpha,\beta}
        \left[
        \Delta_{\alpha\gamma}^*
        W_{\alpha\gamma}W_{\gamma\beta}
        -
        W_{\alpha\gamma}
        \Delta_{\gamma\beta}^*
        W_{\gamma\beta}
        \right]
         \\
        &=
        \sum_{\gamma\neq\alpha,\beta}
        \left(
        \lambda_{\alpha}^{*}
        +
        \lambda_{\beta}^{*}
        -
        2\lambda_{\gamma}^{*}
        \right)
        W_{\alpha\gamma}W_{\gamma\beta}.
    \end{aligned}
\end{equation}

Combining Eqs.~\eqref{eq:supp_offdiag_decay_term} and \eqref{eq:supp_offdiag_quadratic_term} leads to the flow equation
\begin{equation}\label{eq:supp_W_flow}
    \frac{\dif W_{\alpha\beta}}{\dif \ell}
    =
    -
    |\Delta_{\alpha\beta}|^2W_{\alpha\beta}
    +
    \sum_{\gamma\neq\alpha,\beta}
    \left(
    \lambda_{\alpha}^{*}
    +
    \lambda_{\beta}^{*}
    -
    2\lambda_{\gamma}^{*}
    \right)
    W_{\alpha\gamma}W_{\gamma\beta},
    \quad
    \alpha\neq\beta
\end{equation}
for the off-diagonal components.
Together, Eqs.~\eqref{eq:supp_lambda_flow} and \eqref{eq:supp_W_flow} constitute the single-particle non-Hermitian Wegner-flow equations, which are exact at the one-particle level.

To determine the asymptotic diagonal form, we distinguish the realization-specific flow from its statistical average.
In the nonresonant sector, the diagonal flow is of order $O(W^2)$, which implies $\Delta_{\alpha\beta}(\ell)=\Delta_{\alpha\beta}(0)+O(W^2)$.
Retaining terms of $O(W)$ in Eq.~\eqref{eq:supp_W_flow} yields
\begin{equation}
    \frac{\dif W_{\alpha\beta}^{(1)}}{\dif \ell}
    =
    -
    |\Delta_{\alpha\beta}(0)|^2
    W_{\alpha\beta}^{(1)},
\end{equation}
whose solution is given by
\begin{equation}
    \label{eq:supp_W_leading_solution}
    W_{\alpha\beta}^{(1)}(\ell)
    =
    W_{\alpha\beta}(0)
    \mathrm{e}^{-|\Delta_{\alpha\beta}(0)|^2\ell}.
\end{equation}
The nonlinear term generates an $O(W^2)$ correction to the off-diagonal vertex, affecting the diagonal normal form only at $O(W^3)$ and higher orders.
Equation~\eqref{eq:supp_W_leading_solution} therefore establishes the linear stability of the nonresonant diagonal manifold, which suffices to determine the diagonal generator through $O(W^2)$.

The statistical closure is obtained directly from the Gaussian, white-in-time velocity field.
Because $W_{\alpha\beta}$ is linear in the zero-mean velocity, $\langle W_{\alpha\beta}\rangle=0$, while homogeneity and stationarity require
\begin{equation}
    \left\langle
    W_{\alpha\gamma}(0)W_{\gamma\beta}(0)
    \right\rangle
    \propto
    \delta_{\alpha\beta},
\end{equation}
where $\delta_{\alpha\beta}$ represents momentum and frequency conservation in the composite index.
Thus, at bare second order, the averaged quadratic source in the off-diagonal flow vanishes for $\alpha\neq\beta$, whereas the diagonal loop $\alpha\rightarrow\gamma\rightarrow\alpha$ survives.
The latter represents two random scattering events with opposite momentum and frequency transfers, yielding the deterministic self-energy and pair interaction.
Although higher even-order moments do not vanish, Gaussian white-noise averaging reduces them to repeated applications of this time-local, second-order generator instead of introducing new irreducible statistical vertices.
Accordingly, the averaged TK generator is exact without requiring a weak-noise assumption, whereas Eq.~\eqref{eq:supp_W_leading_solution} remains the leading realization-specific flow near the diagonal manifold.

Next, we calculate the leading nonvanishing displacement of the diagonal fixed point.
Since the diagonal flow in Eq.~\eqref{eq:supp_lambda_flow} is quadratic in $W$, substituting the leading trajectory from Eq.~\eqref{eq:supp_W_leading_solution} determines $\lambda_{\alpha}(\infty)$ to second order.
Corrections to the off-diagonal trajectory modify the diagonal eigenvalue only at third order.
We thus obtain
\begin{equation}\label{S-eq:dlamalphadl}
    \frac{\dif \lambda_{\alpha}}{\dif \ell}
    =
    2
    \sum_{\gamma\neq\alpha}
    \Delta_{\alpha\gamma}^{*}(0)
    W_{\alpha\gamma}(0)
    W_{\gamma\alpha}(0)
    \mathrm{e}^{
        -2|\Delta_{\alpha\gamma}(0)|^2\ell
    }.
\end{equation}
Integrating Eq.~\eqref{S-eq:dlamalphadl} from $\ell=0$ to $\ell=\infty$ yields
\begin{equation}
    \begin{aligned}
        \lambda_{\alpha}(\infty)-\lambda_{\alpha}(0)
        &=
        2
        \sum_{\gamma\neq\alpha}
        \Delta_{\alpha\gamma}^{*}(0)
        W_{\alpha\gamma}(0)
        W_{\gamma\alpha}(0)
        \int_0^{\infty}
        \dif\ell\,
        \mathrm{e}^{
            -2|\Delta_{\alpha\gamma}(0)|^2\ell
        }
         \\
        &=
        2
        \sum_{\gamma\neq\alpha}
        \Delta_{\alpha\gamma}^{*}(0)
        W_{\alpha\gamma}(0)
        W_{\gamma\alpha}(0)
        \frac{1}{
            2|\Delta_{\alpha\gamma}(0)|^2
        }
         \\
        &=
        \sum_{\gamma\neq\alpha}
        \frac{
            W_{\alpha\gamma}(0)
            W_{\gamma\alpha}(0)
        }{
            \Delta_{\alpha\gamma}(0)
        }.
    \end{aligned}
\end{equation}
The diagonal normal form through second order is therefore
\begin{equation}
    \label{eq:supp_second_order_lambda}
    \varLambda_{\alpha}
    :=
    \lambda_{\alpha}(\infty)
    =
    \lambda_{\alpha}
    +
    \sum_{\gamma\neq\alpha}
    \frac{
        W_{\alpha\gamma}
        W_{\gamma\alpha}
    }{
        \Delta_{\alpha\gamma}
    },
\end{equation}
where all quantities on the right-hand side are evaluated at $\ell=0$.
The asymptotic suppression of $W$ identifies the diagonal fixed manifold, whereas Eq.~\eqref{eq:supp_second_order_lambda} gives its location in terms of the bare operator through second order.
The Gaussian white-in-time statistics enter subsequently through the ensemble average of the induced second-order vertex.
To this order, the effective diagonal operator is
\begin{equation}
    \mathscr H_{\mathrm{eff}}
    =
    \sum_{\alpha}
    \varLambda_{\alpha}
    |\alpha\rangle
    \langle\alpha|,
    \label{eq:supp_H_eff_composite}
\end{equation}
where
$|\alpha\rangle:=|\vec k,\omega\rangle$.

For passive-scalar turbulence, substituting
Eqs.~\eqref{S-eq:lambda_alpha},
\eqref{S-eq:W_alphabeta}, and
\eqref{eq:supp_Delta_def}
into Eq.~\eqref{eq:supp_second_order_lambda} gives
\begin{equation}
    \label{eq:supp_Lambda_with_Sigma}
    \varLambda_{\vec k,\omega}
    =
    \lambda_{\vec k,\omega}
    +
    \varSigma_{\vec k,\omega},
\end{equation}
where the leading one-particle self-energy is
\begin{equation}
    \label{eq:supp_single_particle_self_energy}
    \varSigma_{\vec k,\omega}
    =
    \sum_{\vec p\neq\vec k}
    \int_{-\infty}^{\infty}
    \frac{\dif\omega'}{2\pi}
    \frac{
        |
        \vec k\cdot
        \hat{\vec v}_{
            \vec k-\vec p,
            \omega-\omega'
        }
        |^2
    }{
        (\omega-\omega')
        -
        (\vec k-\vec p)
        \cdot
        (\vec U+\vec V)
        +
        \mathrm{i}\varkappa(k^2-p^2)
    }.
\end{equation}

\subsection{$n$-particle master operator from the Wegner normal form}\label{sec:supp_n_particle_master_wegner}

Starting from the $n$-particle Wegner normal form, we derive the closed equal-time $n$-point operator.
We define the $n$-point spatial correlation function
\begin{equation}
    C_n(\vec x_1,\cdots,\vec x_n;t)
    =
    \bigg\langle
    \prod_{j=1}^n \theta(\vec x_j,t)
    \bigg\rangle,
\end{equation}
which admits the Fourier representation
\begin{equation}\label{eq:supp_Cn_fourier_H}
    C_n(\vec x_1,\cdots,\vec x_n;t)
    =
    \sum_{\vec k_1,\cdots,\vec k_n}
    \mathrm{e}^{\mathrm{i}\sum_{r=1}^n\vec k_r\cdot\vec x_r}
    C^{(n)}_{\vec k_1,\cdots,\vec k_n}(t).
\end{equation}
For spatially homogeneous correlations, this expansion is restricted to the sector $\sum_{r=1}^n\vec k_r=\vec 0$.

To simplify the notation, we introduce the compact multi-indices $\boldsymbol\alpha=(\alpha_1,\cdots,\alpha_n)$ and $\alpha_r=(\vec k_r,\omega_r)$.
The $n$-particle lift of Eq.~\eqref{eq:supp_composite_constraint} is given by
\begin{equation}\label{eq:supp_Hn_decomposition}
    \mathscr H^{(n)}
    =
    \mathscr H_{\mathrm d}^{(n)}
    +
    \mathscr H_{\mathrm{od}}^{(n)},
    \quad
    \mathscr H_{\mathrm{od}}^{(n)}
    =
    \sum_{r=1}^n
    \mathscr H_{\mathrm{od}}^{[r]} ,
\end{equation}
where the superscript ``$(n)$'' denotes the $n$-particle sector, and ``$[r]$'' labels the operator acting on the $r$-th scalar leg.
Here, a scalar leg represents an external scalar factor in the correlation function.
In the extended-space notation, the $r$-th scalar leg is labeled by $\alpha_r=(\vec k_r,\omega_r)$.
The diagonal matrix elements are given by
\begin{equation}\label{eq:supp_Hn_diagonal}
    \left(
    \mathscr H_{\mathrm d}^{(n)}
    \right)_{\boldsymbol\alpha\boldsymbol\beta}
    =
    \lambda_{\boldsymbol\alpha}^{(n)}
    \delta_{\boldsymbol\alpha\boldsymbol\beta},
    \quad
    \lambda_{\boldsymbol\alpha}^{(n)}
    =
    \sum_{r=1}^n
    \lambda_{\alpha_r},
\end{equation}
while the single-leg off-diagonal matrix elements read
\begin{equation}\label{eq:supp_Hod_leg_matrix}
    \left(
    \mathscr H_{\mathrm{od}}^{[r]}
    \right)_{\boldsymbol\alpha\boldsymbol\beta}
    =
    W_{\alpha_r\beta_r}
    \prod_{m\ne r}
    \delta_{\alpha_m\beta_m},
\end{equation}
where $\beta_r=(\vec p_r,\omega'_r)$.
Consequently, each off-diagonal Wegner vertex modifies only a single scalar leg.

In the $n$-particle sector, the second-order Wegner correction is generated by a pair of off-diagonal vertices. 
We denote this correction by $\Delta\mathscr H^{(n)}$, yielding the effective Hamiltonian
\begin{equation}\label{eq:supp_Heff_n_definition}
    \mathscr H_{\mathrm{eff}}^{(n)}
    =
    \mathscr H_{\mathrm d}^{(n)}
    +
    \Delta\mathscr H^{(n)}
    +
    O\!\Big[
    \left(
    \mathscr H_{\mathrm{od}}^{(n)}
    \right)^3
    \Big].
\end{equation}
For the white-in-time TK model, the frequency denominator in the Wegner normal form admits the equivalent retarded Markov representation
\begin{equation}\label{eq:supp_H_white_kernel}
    \Delta\mathscr H^{(n)}
    =
    \mathrm{i}
    \int_0^\infty
    \left\langle
    \mathscr H_{\mathrm{od}}^{(n)}(t)
    \mathscr H_{\mathrm{od}}^{(n)}(t-\tau)
    \right\rangle_{\vec v}
    \dif\tau .
\end{equation}
The prefactor $\mathrm{i}$ follows from the definition $\mathscr H=\mathrm{i}\partial_t\mathbb I-\mathrm{i}L$, and we refrain from introducing a separate $L$-space notation.
According to Eq.~\eqref{eq:supp_Hn_decomposition}, this second-order correction decomposes to
\begin{equation}\label{eq:supp_DeltaH_rs_definition}
    \Delta\mathscr H^{(n)}
    =
    \sum_{r,s=1}^{n}
    \Delta\mathscr H_{rs}^{(n)},
    \quad
    \Delta\mathscr H_{rs}^{(n)}
    =
    \mathrm{i}
    \int_0^\infty
    \left\langle
    \mathscr H_{\mathrm{od}}^{[r]}(t)
    \mathscr H_{\mathrm{od}}^{[s]}(t-\tau)
    \right\rangle_{\vec v}
    \dif\tau .
\end{equation}
The terms with $r=s$ represent self-contractions, while those with $r\ne s$ correspond to cross-contractions induced by the same velocity field.

For a test amplitude $F_{\vec k_1,\cdots,\vec k_n}$, the action of the time-domain off-diagonal vertex associated with Eq.~\eqref{eq:supp_Hod_leg_matrix} is given by
\begin{equation}\label{eq:supp_Hod_leg_action}
    \left(
    \mathscr H_{\mathrm{od}}^{[r]}(t)F
    \right)_{\vec k_1,\cdots,\vec k_n}
    =
    -
    \sum_{\vec q}
    k_r^i
    \hat v^i_{\vec q}(t)
    F_{\vec k_1,\cdots,\vec k_r-\vec q,\cdots,\vec k_n}.
\end{equation}
The successive application of two such vertices yields
\begin{equation}\label{eq:supp_Hod_two_leg_action}
    \left(
    \mathscr H_{\mathrm{od}}^{[r]}(t)
    \mathscr H_{\mathrm{od}}^{[s]}(t-\tau)F
    \right)_{\vec k_1,\cdots,\vec k_n}
    =
    \sum_{\vec q,\vec q'}
    k_r^i
    \left(
    k_s^j-\delta_{rs}q^j
    \right)
    \hat v^i_{\vec q}(t)
    \hat v^j_{\vec q'}(t-\tau)
    F_{\boldsymbol k^{rs}(\vec q,\vec q')},
\end{equation}
with the shifted multi-index
\begin{equation}\label{eq:supp_k_shift_two_vertex}
    \left[
    \boldsymbol k^{rs}(\vec q,\vec q')
    \right]_m
    =
    \vec k_m
    -
    \delta_{mr}\vec q
    -
    \delta_{ms}\vec q' .
\end{equation}
The factor $k_s^j-\delta_{rs}q^j$ reflects the fact that, when $r=s$, the second vertex acts on the momentum after the shift induced by the first vertex.

Substituting the velocity covariance
\begin{equation}
    \langle
    \hat v^i_{\vec q}(t)
    \hat v^j_{\vec q'}(t')
    \rangle_{\vec v}
    =
    2\hat D_{ij}(\vec q)
    \delta_{\vec q+\vec q',\vec 0}
    \delta(t-t')
\end{equation}
and using the incompressibility constraints $q^i\hat D_{ij}(\vec q)=\hat D_{ij}(\vec q)q^j=0$ in Eqs.~\eqref{eq:supp_DeltaH_rs_definition} to \eqref{eq:supp_k_shift_two_vertex}, we obtain
\begin{equation}\label{eq:supp_DeltaH_rs_raw}
    \left(
    \Delta\mathscr H_{rs}^{(n)}F
    \right)_{\boldsymbol k}
    =
    \mathrm{i}
    \sum_{\vec q}
    k_r^i
    \left(
    k_s^j-\delta_{rs}q^j
    \right)
    \hat D_{ij}(\vec q)
    F_{\boldsymbol k^{rs}(\vec q)},
\end{equation}
where $[\boldsymbol k^{rs}(\vec q)]_m=\vec k_m-\delta_{mr}\vec q + \delta_{ms}\vec q$.
Note that the term proportional to $q^j$ vanishes by incompressibility.
Summing over all scalar legs gives
\begin{equation}\label{eq:supp_DeltaH_spectral}
    \left(
    \Delta\mathscr H^{(n)}F
    \right)_{\boldsymbol k}
    =
    \mathrm{i}
    \sum_{r,s=1}^n
    \sum_{\vec q}
    k_r^i k_s^j
    \hat D_{ij}(\vec q)
    F_{\boldsymbol k^{rs}(\vec q)}.
\end{equation}

Combining the deterministic diagonal part with Eq.~\eqref{eq:supp_DeltaH_spectral} yields the effective equal-time $n$-particle operator
\begin{equation}\label{eq:supp_H_eff_spectral}
    \left(
    \mathscr H_{\mathrm{eff}}^{(n)}F
    \right)_{\boldsymbol k}
    =
    \bigg[
    \mathrm{i}\partial_t
    -
    (\vec U+\vec V)\cdot
    \sum_{r=1}^n\vec k_r
    +
    \mathrm{i}\varkappa
    \sum_{r=1}^n k_r^2
    \bigg]
    F_{\boldsymbol k}
    +
    \mathrm{i}
    \sum_{r,s=1}^n
    \sum_{\vec q}
    k_r^i k_s^j
    \hat D_{ij}(\vec q)
    F_{\boldsymbol k^{rs}(\vec q)}
\end{equation}
in Fourier space.
Substituting $D_{ij}(\vec x)=\sum_{\vec q}\hat D_{ij}(\vec q)\mathrm{e}^{\mathrm{i}\vec q\cdot\vec x}$ and taking inverse Fourier transform of Eq.~\eqref{eq:supp_H_eff_spectral} yield the explicit coordinate-space operator
\begin{equation}\label{eq:supp_H_eff_spatial}
    \mathscr H_{\mathrm{eff}}^{(n)}
    =
    \mathrm{i}\partial_t
    +
    \mathrm{i}(\vec U+\vec V)\cdot
    \sum_{r=1}^n\bn_{\vec x_r}
    -
    \mathrm{i}\varkappa
    \sum_{r=1}^n\nabla_{\vec x_r}^2
    -
    \mathrm{i}
    \sum_{r,s=1}^n
    D_{ij}(\vec x_r-\vec x_s)
    \partial_{x_r^i}\partial_{x_s^j}.
\end{equation}
Consequently, the equal-time correlation satisfies
\begin{equation}
    \mathscr H_{\mathrm{eff}}^{(n)}C_n
    =
    \mathrm{i}\varPhi_n .
    \label{eq:supp_H_eff_Cn_equation}
\end{equation}
For Gaussian white scalar forcing with covariance $\chi(\vec x-\vec y)\delta(t-t')$, the source term is
\begin{equation}\label{eq:supp_force_hierarchy_H}
    \varPhi_n
    =
    \sum_{1\le r<s\le n}
    \chi(\vec x_r-\vec x_s)
    C_{n-2}
    \left(
    \{\vec x_m\}_{m\ne r,s};t
    \right).
\end{equation}

For translationally invariant correlations, the center-of-mass derivative annihilates $C_n$, namely $\sum_{r=1}^n\nabla_{\vec x_r}C_n=0$.
Hence, the uniform-advection term in Eq.~\eqref{eq:supp_H_eff_spatial} vanishes on the homogeneous sector.
The same projection also eliminates the zero-separation part of the velocity covariance.
Introducing the structure tensor
\begin{equation}\label{eq:supp_dij_def_H}
    d_{ij}(\vec x)
    :=
    D_{ij}(\vec 0)-D_{ij}(\vec x)
    = D_1 r^\xi
	\left[
	(d-1+\xi)\delta_{ij}
	-
	\xi \frac{r_i r_j}{r^2}
	\right],
\end{equation}
we obtain
\begin{equation}\label{eq:supp_D_to_d_decomposition}
    \sum_{r,s=1}^n
    D_{ij}(\vec x_r-\vec x_s)
    \partial_{x_r^i}\partial_{x_s^j}
    =
    D_{ij}(\vec 0)
    \bigg(
    \sum_{r=1}^n
    \partial_{x_r^i}
    \bigg)
    \bigg(
    \sum_{s=1}^n
    \partial_{x_s^j}
    \bigg)
    -
    \sum_{r,s=1}^n
    d_{ij}(\vec x_r-\vec x_s)
    \partial_{x_r^i}\partial_{x_s^j}.
\end{equation}
The first term in Eq.~\eqref{eq:supp_D_to_d_decomposition} acts only on the center-of-mass coordinate and therefore vanishes on translationally invariant correlations.
Consequently, on the homogeneous sector, we obtain
\begin{equation}\label{eq:supp_D_to_d_H}
    \sum_{r,s=1}^n
    D_{ij}(\vec x_r-\vec x_s)
    \partial_{x_r^i}\partial_{x_s^j}
    =
    -
    \sum_{r,s=1}^n
    d_{ij}(\vec x_r-\vec x_s)
    \partial_{x_r^i}\partial_{x_s^j}.
\end{equation}
Thus, $d_{ij}$ isolates the relative velocity statistics acting on separations, whereas the discarded $D_{ij}(\vec 0)$ term represents the common sweeping contribution.

Following the homogeneous projection, the full relative-coordinate effective operator retains the time derivative and is expressed as
\begin{equation}\label{eq:supp_H_relative_full}
    \mathscr H_{\mathrm{eff}}^{(n)}
    =
    \mathrm{i}\partial_t
    -
    \mathrm{i}\varkappa
    \sum_{r=1}^n\nabla_{\vec x_r}^2
    +
    \mathrm{i}
    \sum_{r,s=1}^n
    d_{ij}(\vec x_r-\vec x_s)
    \partial_{x_r^i}\partial_{x_s^j}.
\end{equation}
Consequently, the projected $n$-point equation reads
\begin{equation}\label{eq:supp_H_relative_full_equation}
    \mathscr H_{\mathrm{eff}}^{(n)} C_n
    =
    \mathrm{i}\varPhi_n .
\end{equation}
For stationary equal-time correlations, the condition $\partial_t C_n=0$ holds.
The spatial relative operator associated with the steady hierarchy is then given by
\begin{equation}\label{eq:supp_H_relative}
    \mathscr H_{\mathrm{eff}}^{(n)}
    =
    -\mathrm{i}\varkappa
    \sum_{r=1}^n\nabla_{\vec x_r}^2
    +
    \mathrm{i}
    \sum_{r,s=1}^n
    d_{ij}(\vec x_r-\vec x_s)
    \partial_{x_r^i}\partial_{x_s^j}.
\end{equation}

Within the inertial range, the forcing contribution $\varPhi_n$ vanishes locally, and molecular diffusion is subleading away from the dissipative cutoff.
Consequently, the homogeneous zero modes satisfy
\begin{equation}\label{eq:supp_H_zero_mode}
    \mathscr H_{\mathrm{eff}}^{(n)}
    Z_n(\vec x_1,\cdots,\vec x_n)=0.
\end{equation}
Equivalently, removing the overall factor of $-\mathrm{i}$ yields
\begin{equation}\label{eq:supp_zero_mode_real_operator}
    \bigg[
    \varkappa
    \sum_{r=1}^n\nabla_{\vec x_r}^2
    -
    \sum_{r,s=1}^n
    d_{ij}(\vec x_r-\vec x_s)
    \partial_{x_r^i}\partial_{x_s^j}
    \bigg]
    Z_n=0.
\end{equation}

\subsection{Eddy diffusivity and the scalar Boussinesq relation}
We now extract the mean transport law from the one-particle sector of the Wegner normal form.
Let $\varTheta(\boldsymbol{x},t) = \langle \theta(\boldsymbol{x},t) \rangle_{\vec{v},f}$ denote the ensemble-averaged scalar, where the sweeping velocity $\boldsymbol{V}$ is frozen.
This conditional averaging decouples the diffusion induced by the small-scale velocity $\boldsymbol{v}$ from sweeping decorrelation.
Since the forcing has zero mean, it does not contribute to the one-point equation.

For $n=1$, Eq.~\eqref{eq:supp_DeltaH_spectral} reduces to the single self-contraction $r=s=1$. 
Using $[\boldsymbol{k}^{11}(\boldsymbol{q})]_1 =\boldsymbol{k}-\boldsymbol{q}+\boldsymbol{q} =\boldsymbol{k}$, the Wegner correction acting on a test amplitude simplifies to
\begin{equation}
    \left(\Delta\mathscr{H}^{(1)}F\right)_{\boldsymbol{k}}
    =
    \mathrm{i}
    \sum_{\boldsymbol{q}}
    k^i k^j
    \hat{D}_{ij}(\boldsymbol{q})
    F_{\boldsymbol{k}}
    =
    \mathrm{i}
    k^i k^j D_{ij}(\boldsymbol{0})
    F_{\boldsymbol{k}}
    = \ii D_0 k^2 F_{\boldsymbol{k}}.
\end{equation}
Consequently, the one-particle Wegner correction is diagonal in wave-vector space. 
Furthermore, because Eq.~\eqref{eq:supp_H_white_kernel} is local and stationary in time, its extended-space matrix elements take the form
\begin{equation}
    \left(\Delta\mathscr{H}^{(1)}\right)_
    {\boldsymbol{k}\omega,\boldsymbol{p}\omega'}
    =
    2\pi
    \delta_{\boldsymbol{k}\boldsymbol{p}}
    \delta(\omega-\omega')
    \mathrm{i} D_0 k^2.
\end{equation}
By comparing this expression with the diagonal normal form in Eqs.~\eqref{eq:supp_H_eff_composite} and \eqref{eq:supp_Lambda_with_Sigma}, we identify the ensemble-averaged one-particle self-energy as
\begin{equation}
    \varSigma_{\boldsymbol{k},\omega}
    =
    \mathrm{i} D_0 k^2.
\end{equation}
The diagonal eigenvalue then becomes
\begin{equation}
    \varLambda_{\boldsymbol{k},\omega}
    =
    \omega
    -
    \boldsymbol{k}\cdot(\boldsymbol{U}+\boldsymbol{V})
    +
    \mathrm{i}(\varkappa+D_0)k^2 .
\end{equation}
For $n=1$, the forcing sum in Eq.~\eqref{eq:supp_force_hierarchy_H} is empty. 
Taking the inverse Fourier transform of the algebraic constraint then yields
\begin{equation}\label{S-eq:S-eq:Theta_eq_0}
    \left[
        \partial_t
        +
        (\boldsymbol{U}+\boldsymbol{V})\cdot\bn
        -
        (\varkappa+D_0)\nabla^2
    \right]\varTheta
    =0 .
\end{equation}
Thus, the small-scale turbulent velocity yields the eddy and effective diffusivities
\begin{equation}
    D_{\mathrm{eddy}}=D_0,
    \quad
    D_{\mathrm{eff}}=\varkappa+D_0 .
\end{equation}

To identify the associated turbulent flux, we decompose the scalar field as $\theta=\varTheta+\theta'$ and define the flux component as
\begin{equation}
    \mathcal{J}_i
    :=
    \left\langle v_i\theta'\right\rangle_{\boldsymbol{v},f}
    =
    \left\langle v_i\theta\right\rangle_{\boldsymbol{v},f}.
\end{equation}
Averaging Eq.~\eqref{eq:supp_scalar_pde} and using the incompressibility condition $\partial_i v_i=0$ yields
\begin{equation}\label{S-eq:Theta_eq_J_i}
    \left[
        \partial_t
        +
        (\boldsymbol{U}+\boldsymbol{V})\cdot\bn
        -
        \varkappa\nabla^2
    \right]\varTheta
    =
    -\partial_i\mathcal{J}_i .
\end{equation}
Comparing Eq.~\eqref{S-eq:Theta_eq_J_i} with Eq.~\eqref{S-eq:S-eq:Theta_eq_0}, we obtain
\begin{equation}
    -\partial_i\mathcal{J}_i
    =
    D_{ij}(\boldsymbol{0})\partial_i\partial_j\Theta .
\end{equation}
Since the Wegner vertex generates only longitudinal contribution, transforming back to physical space yields
\begin{equation}
    \mathcal{J}_i
    =
    -D_{ij}(\boldsymbol{0})\partial_j\varTheta
    =
    -D_0\partial_i\varTheta.
\end{equation}
This expression corresponds to the scalar Boussinesq relation. 
It is therefore a direct consequence of the diagonal Wegner normal form rather than an independent closure assumption. 
This local form is exact for the homogeneous, Gaussian, and white-in-time TK ensemble. 
Conversely, a finite velocity correlation time introduce a memory kernel and a frequency-dependent turbulent diffusivity~\cite{Wang2026_Space}.

\section{Unequal-time correlations from the Wegner normal form}\label{sec:supp_unequal_time_correlations}
We derive the unequal-time hierarchy using the Wegner normal form.
Related studies of two-time and spatial-temporal passive-scalar correlations are given in Refs.~\cite{Gorbunova2021_Eulerian, Yang2023_Space}.
In contrast to the equal-time case, the homogeneous projection does not generally remove the common sweeping contribution, because only scalar factors with observation times in the future remain dynamically active between consecutive observation times.
These active factors may carry a nonzero total wave vector relative to the factors already observed at earlier times.
Consequently, their center-of-mass sector undergoes mean advection, random sweeping~\cite{Kraichnan1964_Kolmogorov, Tennekes1975_Eulerian}, and diffusion induced by $D_{ij}(\vec 0)$.
The unequal-time normal form must therefore retain this center-of-mass sector before separating the relative-coordinate dynamics.
The resulting formulation involves a time-ordered sequence of active-set operators, rather than a single equal-time $n$-particle operator, and separates the Eulerian sweeping and center-diffusion factors from the relative relaxation modes that encode the dynamic zero-mode structure.

\subsection{Pair vertex generated by the Wegner flow}
Before applying the Wegner flow, we specify the multi-time object and its Fourier representation.
Consider $M$ ordered times $t_1<t_2<\cdots<t_M$.
The $m$-th time slice contains $n_m$ scalar fields, whose spatial arguments are denoted by
\begin{equation}
	X_m
	=
	(\vec x_{m,1},\cdots,\vec x_{m,n_m}),
	\quad
	n
	=
	\sum_{m=1}^{M}n_m .
	\label{eq:supp_time_slice_coordinates}
\end{equation}
Here, $n$ denotes the total order of the correlation.
We define the multi-time Eulerian correlation as
\begin{equation}\label{eq:supp_multi_time_C_def}
	C_{\boldsymbol n}^{\mathrm E}
	(X_M,t_M;\cdots;X_1,t_1)
	=
	\left\langle
	\prod_{m=1}^{M}
	\prod_{r=1}^{n_m}
	\theta(\vec x_{m,r},t_m)
	\right\rangle,
\end{equation}
where $\boldsymbol n=(n_M,\cdots,n_1)$ specifies the number of fields on each time slice, and the superscript ``$\mathrm E$'' denotes an Eulerian correlation.
Taking the Fourier expansion of Eq.~\eqref{eq:supp_multi_time_C_def} yields
\begin{equation}
	C_{\boldsymbol n}^{\mathrm E}
	=
	\sum_{\{\vec k_{m,r}\}}
	\exp\left(
	\ii
	\sum_{m=1}^{M}
	\sum_{r=1}^{n_m}
	\vec k_{m,r}\cdot\vec x_{m,r}
	\right)
	C_{\{\vec k_{m,r}\}}^{(\boldsymbol n)}
	(t_M,\cdots,t_1).
	\label{eq:supp_multi_time_fourier}
\end{equation}
For homogeneous correlations, the Fourier amplitudes are restricted to the total-momentum sector
\begin{equation}
	\sum_{m=1}^{M}
	\sum_{r=1}^{n_m}
	\vec k_{m,r}
	=
	\vec 0 .
	\label{eq:supp_multi_time_total_momentum}
\end{equation}
This condition applies to the entire set of external wave vectors and does not require the wave vectors on each individual time slice to sum to zero.
The set $\{\vec k_{m,r}\}$ denotes all wave vectors carried by the external scalar legs.

We next identify the effective interaction generated by eliminating the off-diagonal scattering vertices.
For a scalar leg labelled by the pair $(m,r)$, we introduce the composite index $\alpha_{m,r}=(\vec k_{m,r},\omega_{m,r})$.
The collection of all external indices is denoted by $\boldsymbol\alpha=\{\alpha_{m,r}\}_{1\le m\le M,\;1\le r\le n_m}$, which gives the unequal-time counterpart of the $n$-particle multi-index in Eq.~\eqref{eq:supp_Hn_decomposition}.
Accordingly, the $n$-particle lifted operator reads
\begin{equation}
	\mathscr H^{(n)}
	=
	\mathscr H_{\mathrm d}^{(n)}
	+
	\mathscr H_{\mathrm{od}}^{(n)},
	\quad
	\mathscr H_{\mathrm{od}}^{(n)}
	=
	\sum_{m=1}^{M}
	\sum_{r=1}^{n_m}
	\mathscr H_{\mathrm{od}}^{[m,r]} .
	\label{eq:supp_multi_time_Hn}
\end{equation}
The operator $\mathscr H_{\mathrm{od}}^{[m,r]}$ acts only on the scalar leg $(m,r)$.
Its action on a test amplitude $F_{\boldsymbol\alpha}$ is given by
\begin{equation}
	\left(
	\mathscr H_{\mathrm{od}}^{[m,r]}F
	\right)_{\boldsymbol\alpha}
	=
	-
	\sum_{\vec q}
	\int_{-\infty}^{\infty}
	\frac{\dif\omega'}{2\pi}\,
	k_{m,r}^{i}
	\hat v_{\vec q,\omega'}^{i}
	F(\cdots,
	\vec k_{m,r}-\vec q,
	\omega_{m,r}-\omega',
	\cdots),
	\label{eq:supp_multi_time_Hod_action}
\end{equation}
where $\omega'$ represents the frequency transferred by the velocity mode.
Thus, a single off-diagonal Wegner vertex changes both the wave vector and the frequency of one scalar leg.

The non-Hermitian Wegner flow suppresses these off-diagonal vertices.
At second order in $\mathscr H_{\mathrm{od}}^{(n)}$, the effective interaction is generated by two virtual scattering events.
For two scalar legs labelled by $(m,r)$ and $(l,s)$, we define
\begin{equation}
	\Delta\mathscr H_{m r,l s}^{(n)}
	=
	\ii
	\int_0^\infty
	\left\langle
	\mathscr H_{\mathrm{od}}^{[m,r]}(t)
	\mathscr H_{\mathrm{od}}^{[l,s]}(t-\tau)
	\right\rangle_{\vec v}
	\dif\tau,
	\label{eq:supp_multi_time_DeltaH_malb_def}
\end{equation}
which is the unequal-time counterpart of Eq.~\eqref{eq:supp_DeltaH_rs_definition}.
We next write the action of a single off-diagonal vertex in frequency space.
For the scalar leg labelled by $(m,r)$, Eq.~\eqref{eq:supp_multi_time_Hod_action} yields
\begin{equation}
	\left(
	\mathscr H_{\mathrm{od}}^{[m,r]}F
	\right)_{\boldsymbol\alpha}
	=
	-
	\sum_{\vec q}
	\int_{-\infty}^{\infty}
	\frac{\dif\omega}{2\pi}\,
	k_{m,r}^{i}
	\hat v_{\vec q,\omega}^{i}
	F\!\left(
	\boldsymbol\alpha^{m r}(\vec q,\omega)
	\right),
	\label{eq:supp_single_vertex_ma}
\end{equation}
with the shifted multi-index
\begin{equation}
	\alpha_{j, k}^{m r}(\vec q,\omega)
	=
	\left(
	\vec k_{j, k}
	-
	\delta_{j m}\delta_{k r}\vec q,
	\,
	\omega_{j, k}
	-
	\delta_{j m}\delta_{k r}\omega
	\right), \quad j\in\{1,2,\cdots,M\},~k\in\{1,2,\cdots,n_j\}.
	\label{eq:supp_single_shift_alpha_ma}
\end{equation}
Thus, the vertex changes only the wave vector and the frequency of the scalar leg on which it acts.
Here, $\vec k_{j, k}$ and $\omega_{j, k}$ denote the wave vector and frequency carried by the $k$-th scalar factor on the $j$-th time slice, respectively.

We next apply two such vertices.
Using Eq.~\eqref{eq:supp_single_vertex_ma} twice gives
\begin{equation}
	\left(
	\mathscr H_{\mathrm{od}}^{[m,r]}
	\mathscr H_{\mathrm{od}}^{[l,s]}F
	\right)_{\boldsymbol\alpha}
	=
	\sum_{\vec q,\vec q'}
	\int_{-\infty}^{\infty}
	\frac{\dif\omega}{2\pi}
	\int_{-\infty}^{\infty}
	\frac{\dif\omega'}{2\pi}\,
	\left(
	k_{m,r}^{i}
	-
	\delta_{ml}\delta_{rs}q'^{\,i}
	\right)
	k_{l,s}^{j}
	\hat v_{\vec q,\omega}^{i}
	\hat v_{\vec q',\omega'}^{j}
	F\!\left(
	\boldsymbol\alpha^{m r,l s}
	(\vec q,\vec q',\omega,\omega')
	\right),
	\label{eq:supp_two_vertex_before_average}
\end{equation}
with the shifted multi-index
\begin{equation}
	\alpha_{j, k}^{m r,l s}
	(\vec q,\vec q',\omega,\omega')
	=
	\Big(
	\vec k_{j, k}
	-
	\delta_{j m}\delta_{k r}\vec q
	-
	\delta_{j l}\delta_{k s}\vec q',
	\omega_{j, k}
	-
	\delta_{j m}\delta_{k r}\omega
	-
	\delta_{j l}\delta_{k s}\omega'
	\Big).
	\label{eq:supp_two_vertex_shift_alpha_before_average}
\end{equation}
The factor $k_{m,r}^{i} - \delta_{ml}\delta_{rs}q'^{\,i}$ arises only when the two vertices act on the same scalar leg.
In this case, the second vertex acts on a leg whose momentum has already been shifted by the first vertex.

We now average over the white-in-time Gaussian velocity field.
Substituting the velocity covariance
\begin{equation}
    \langle\hat v_{\vec q,\omega}^{i}\hat v_{\vec q',\omega'}^{j}\rangle = 4\pi \hat D_{ij}(\vec q) \delta_{\vec q+\vec q',\vec 0} \delta(\omega+\omega')
\end{equation}
into Eq.~\eqref{eq:supp_two_vertex_before_average} yields
\begin{equation}
	\left\langle
	\left(
	\mathscr H_{\mathrm{od}}^{[m,r]}
	\mathscr H_{\mathrm{od}}^{[l,s]}F
	\right)_{\boldsymbol\alpha}
	\right\rangle
	=
	2
	\sum_{\vec q}
	\int_{-\infty}^{\infty}
	\frac{\dif\omega}{2\pi}\,
	\left(
	k_{m,r}^{i}
	+
	\delta_{ml}\delta_{rs}q^{i}
	\right)
	k_{l,s}^{j}
	\hat D_{ij}(\vec q)
	F\!\left(
	\boldsymbol\alpha^{m r,l s}
	(\vec q,-\vec q,\omega,-\omega)
	\right).
	\label{eq:supp_two_vertex_after_average_raw}
\end{equation}
The term proportional to $q^i$ vanishes because incompressibility imposes the condition $q^i\hat D_{ij}(\vec q)=0$.
Thus, we obtain
\begin{equation}
	\left\langle
	\left(
	\mathscr H_{\mathrm{od}}^{[m,r]}
	\mathscr H_{\mathrm{od}}^{[l,s]}F
	\right)_{\boldsymbol\alpha}
	\right\rangle
	=
	2
	\sum_{\vec q}
	\int_{-\infty}^{\infty}
	\frac{\dif\omega}{2\pi}\,
	k_{m,r}^{i}
	k_{l,s}^{j}
	\hat D_{ij}(\vec q)
	F\!\left(
	\boldsymbol\alpha^{m r,l s}
	(\vec q,\omega)
	\right),
	\label{eq:supp_two_vertex_after_average}
\end{equation}
with the reduced shifted multi-index
\begin{equation}
	\alpha_{j, k}^{m r,l s}
	(\vec q,\omega)
	=
	\Big(
	\vec k_{j, k}
	-
	\delta_{j m}\delta_{k r}\vec q
	+
	\delta_{j l}\delta_{k s}\vec q,
	\omega_{j, k}
	-
	\delta_{j m}\delta_{k r}\omega
	+
	\delta_{j l}\delta_{k s}\omega
	\Big).
	\label{eq:supp_multi_time_shifted_alpha_reduced}
\end{equation}
This shifted index conserves the total wave vector and the total frequency of the two scalar factors.

Finally, using the half-delta convention $\int_0^\infty \delta(\tau)\,\dif\tau=\frac{1}{2}$, we obtain
\begin{equation}
	\left(
	\Delta\mathscr H_{m r,l s}^{(n)}F
	\right)_{\boldsymbol\alpha}
	=
	\ii
	\sum_{\vec q}
	\int_{-\infty}^{\infty}
	\frac{\dif\omega}{2\pi}\,
	k_{m,r}^{i}k_{l,s}^{j}
	\hat D_{ij}(\vec q)
	F\big(
	\boldsymbol\alpha^{m r,l s}(\vec q,\omega)
	\big),
	\label{eq:supp_multi_time_pair_vertex_frequency}
\end{equation}
with the shifted multi-index
\begin{equation}
	\alpha_{j,k}^{m r,l s}(\vec q,\omega)
	=
	\Big(
	\vec k_{j,k}
	-
	\delta_{j m}\delta_{kr}\vec q
	+
	\delta_{j l}\delta_{ks}\vec q,
	\omega_{j,k}
	-
	\delta_{j m}\delta_{kr}\omega
	+
	\delta_{j l}\delta_{ks}\omega
	\Big).
	\label{eq:supp_multi_time_shifted_alpha}
\end{equation}
Equation~\eqref{eq:supp_multi_time_pair_vertex_frequency} is the pairwise effective interaction generated by the second-order Wegner normal form, which changes two scalar legs by equal and opposite wave vectors and by equal and opposite frequencies.
Consequently, it preserves both the total momentum and the total frequency of the pair.

\subsection{Time-local interaction and active-set generator}
We then transform the external frequencies back to the external times.
Specifically, integrating over the frequency transferred by the velocity mode yields
\begin{equation}
	\int_{-\infty}^{\infty}
	\frac{\dif\omega}{2\pi}
	\ee^{\ii\omega(t_m-t_l)}
	=
	\delta(t_m-t_l).
	\label{eq:supp_frequency_contact_time}
\end{equation}
Thus, the interaction generated by the Wegner normal form is local in the external times.
This delta function constrains the two scalar legs connected by a velocity mode to the same time slice.
Consequently, all scalar pairs within a given time slice contribute at equal times.
For unequal external times, only those scalar pairs contained in the active time slice contribute.
This contact structure underlies the active-set dynamics.
During a given time interval, only fields with future observation times are propagated.
The Wegner vertex then couples only pairs within this active set.

We next derive the time-domain evolution generated by Eq.~\eqref{eq:supp_multi_time_pair_vertex_frequency}.
Let $A$ denote the active set of scalar legs with coordinates $\{\vec y_r\}_{r\in A}$.
We define the active-set gradient as
\begin{equation}
	\bn_A
	:=
	\sum_{r\in A}
	\bn_{\vec y_r}.
	\label{eq:supp_active_gradient}
\end{equation}
The corresponding Eulerian real-time generator governing the homogeneous part of the correlation, $\partial_t C_{\vec n}^{\mathrm{E, hom}}=\mathcal L_A^{\mathrm E}C_{\vec n}^{\mathrm{E, hom}}$, is
\begin{equation}
	\mathcal L_{A}^{\mathrm E}
	=
	-(\vec U+\vec V)\cdot\bn_A
	+
	\varkappa
	\sum_{r\in A}
	\nabla_{\vec y_r}^{2}
	+
	\sum_{r,s\in A}
	D_{ij}(\vec y_r-\vec y_s)
	\partial_{y_r^i}\partial_{y_s^j}.
	\label{eq:supp_active_L_E}
\end{equation}
Here, $\mathcal L_A^{\mathrm E}$ acts directly in real time.
It is obtained from $\mathscr H_{\mathrm{eff}}^{(n)}$ by removing the overall factor $\ii$ and isolating $\partial_t$.
In the equal-time sector, where $A=\{1,\cdots,n\}$, Eq.~\eqref{eq:supp_active_L_E} reduces to Eq.~\eqref{eq:supp_H_eff_spatial}.

Using $D_{ij}(\vec r) = D_{ij}(\vec 0)-d_{ij}(\vec r)$, we decompose Eq.~\eqref{eq:supp_active_L_E} into center-of-mass and relative contributions as
\begin{equation}
	\mathcal L_A^{\mathrm E}
	=
	-(\vec U+\vec V)\cdot\bn_A
	+
	D_{ij}(\vec 0)\bn_A^i\bn_A^j
	+
	\mathcal M_A,
	\label{eq:supp_active_L_decomposition}
\end{equation}
where the relative-coordinate generator is given by
\begin{equation}
	\mathcal M_A
	=
	\varkappa
	\sum_{r\in A}
	\nabla_{\vec y_r}^{2}
	-
	\sum_{r,s\in A}
	d_{ij}(\vec y_r-\vec y_s)
	\partial_{y_r^i}\partial_{y_s^j}.
	\label{eq:supp_active_M_def}
\end{equation}
The first two terms in Eq.~\eqref{eq:supp_active_L_decomposition} depend exclusively on the total gradient $\bn_A$.
These terms govern the center-of-mass motion of the active set, dropping out only after the projection $\bn_A C=0$ is enforced.
In equal-time homogeneous correlations, this projection is satisfied automatically because all external legs are active.
However, for unequal-time Eulerian correlations, only future legs are active.
As a result, $\bn_A C$ is generally nonvanishing, meaning the center-of-mass terms must be retained.

Then, we derive the ordered multi-time correlation.
For $m=2,\cdots,M$, we define the active set
\begin{equation}
	A_m
	=
	X_m\cup X_{m+1}\cup\cdots\cup X_M
	\label{eq:supp_active_set_Am}
\end{equation}
with the time increment $\Delta t_m=t_m-t_{m-1}$, which contains all scalar factors whose observation times are later than or equal to $t_m$.
Between the two adjacent observation times $t_{m-1}$ and $t_m$, only the scalar factors in $A_m$ remain to be propagated.
The homogeneous contribution contains no scalar-forcing insertion inside this open interval, yielding the autonomous linear equation
\begin{equation}
	\partial_t C
	=
	\mathcal L_{A_m}^{\mathrm E}C,
	\quad
	t_{m-1}<t<t_m .
	\label{eq:supp_interval_linear_evolution}
\end{equation}
The corresponding evolution operator over this interval is the semigroup
\begin{equation}
	C(t_m)
	=
	\ee^{\Delta t_m\mathcal L_{A_m}^{\mathrm E}}
	C(t_{m-1}) .
	\label{eq:supp_interval_semigroup}
\end{equation}
Thus, the full homogeneous multi-time correlation follows from the equal-time boundary value at $t_1$ by successive application of these semigroups across the prescribed observation times.
Consequently, the source-free part of the multi-time correlation is obtained from the time-ordered evolution
\begin{equation}
	C_{\boldsymbol n}^{\mathrm E,\mathrm{sf}}
	=
	\ee^{\Delta t_M\mathcal L_{A_M}^{\mathrm E}}
	\ee^{\Delta t_{M-1}\mathcal L_{A_{M-1}}^{\mathrm E}}
	\cdots
	\ee^{\Delta t_2\mathcal L_{A_2}^{\mathrm E}}
	C_n(X_M,\cdots,X_1; t_1).
	\label{eq:supp_multi_time_hom_solution}
\end{equation}
Here, $C_n$ denotes the equal-time $n$-point correlation evaluated at $(X_M,\cdots,X_1)$, which provides the boundary condition
\begin{equation}
	C_{\boldsymbol n}^{\mathrm E}
	(X_M,t_1^+;\cdots;X_1,t_1)
	=
	C_n(X_M,\cdots,X_1; t_1)
	\label{eq:supp_multi_time_fusion}
\end{equation} 
at the earliest time $t_1$.
The superscript ``$\mathrm{sf}$'' indicates that no scalar-forcing source is inserted within the open intervals $(t_{m-1},t_m)$.
The full correlation $C_{\boldsymbol n}^{\mathrm E}$ is obtained by adding the Duhamel terms generated by scalar-forcing sources inside the open intervals.

\subsection{Sweeping and center-diffusion factors}
We next separate the center-of-mass and relative-coordinate sectors in Fourier space.
For the active set $A_m$, we first define a total wave vector
\begin{equation}
	\vec K_m
	=
	\sum_{i=m}^{M}
	\sum_{j=1}^{n_i}
	\vec k_{i,j},
	\quad
	m=2,\cdots,M.
	\label{eq:supp_active_Km_def}
\end{equation}
In Fourier space, the center-of-mass part of the propagator generated by the first two terms in Eq.~\eqref{eq:supp_active_L_decomposition} contributes
\begin{equation}
	\prod_{m=2}^{M}
	\ee^{-\ii(\vec U+\vec V)\cdot\vec K_m\Delta t_m}
	\ee^{-D_{ij}(\vec 0)K_m^iK_m^j\Delta t_m}
    = \prod_{m=2}^{M}
	\ee^{-\ii(\vec U+\vec V)\cdot\vec K_m\Delta t_m}
	\ee^{-D_0 K_m^2 \Delta t_m}.
	\label{eq:supp_center_factor_given_V}
\end{equation}
Averaging over the Gaussian sweeping velocity $\vec V$ yields
\begin{equation}
	\left\langle
	\prod_{m=2}^{M}
	\ee^{-\ii\vec V\cdot\vec K_m\Delta t_m}
	\right\rangle_{\vec V}
	=
	\exp\bigg(
	-\frac{V_0^2}{2}
	\bigg|
	\sum_{m=2}^{M}
	\vec K_m\Delta t_m
	\bigg|^2 
	\bigg).
	\label{eq:supp_sweeping_average_general}
\end{equation}
Using the identity (proof detailed in Sec.~\ref{sec:proof_supp_telescoping_identity})
\begin{equation}
	\sum_{m=2}^{M}
	\vec K_m\Delta t_m
	=
	\sum_{m=1}^{M}
	\sum_{r=1}^{n_m}
	\vec k_{m,r}t_m,
	\label{eq:supp_telescoping_identity}
\end{equation}
we factorize the Eulerian multi-time correlation as
\begin{equation}
	C_{\{\vec k_{m,r}\}}^{(\boldsymbol n)}
	(t_M,\cdots,t_1)
	=
	\ee^{
	-\ii\vec U\cdot
	\sum_{m=1}^{M}
	\sum_{r=1}^{n_m}
	\vec k_{m,r}t_m
	}
	\ee^{
	-\frac{V_0^2}{2}
	\left|
	\sum_{m=1}^{M}
	\sum_{r=1}^{n_m}
	\vec k_{m,r}t_m
	\right|^2
	}
	\prod_{m=2}^{M}
	\ee^{-D_0 K_m^2 \Delta t_m}
	C_{\{\vec k_{m,r}\}}^{(\boldsymbol n),\mathrm{rel}}
	(t_M,\cdots,t_1),
	\label{eq:supp_multi_time_factorization}
\end{equation}
with the relative part 
\begin{equation}
	C_{\boldsymbol n}^{\mathrm{rel}}
	=
	\ee^{\Delta t_M\mathcal M_{A_M}}
	\ee^{\Delta t_{M-1}\mathcal M_{A_{M-1}}}
	\cdots
	\ee^{\Delta t_2\mathcal M_{A_2}}
	C_n.
	\label{eq:supp_multi_time_relative_part}
\end{equation}
The Fourier coefficients $C_{\{\vec k_{m,r}\}}^{(\boldsymbol n),\mathrm{rel}}$ and the corresponding physical-space correlation $C_{\boldsymbol n}^{\mathrm{rel}}$ are related by the same Fourier convention as in Eq.~\eqref{eq:supp_multi_time_fourier}, namely
\begin{equation}
	C_{\boldsymbol n}^{\mathrm{rel}}
	(X_M,t_M;\cdots;X_1,t_1)
	=
	\sum_{\{\vec k_{m,r}\}}
	\exp\left(
	\ii
	\sum_{m=1}^{M}
	\sum_{r=1}^{n_m}
	\vec k_{m,r}\cdot\vec x_{m,r}
	\right)
	C_{\{\vec k_{m,r}\}}^{(\boldsymbol n),\mathrm{rel}}
	(t_M,\cdots,t_1).
	\label{eq:supp_multi_time_relative_fourier}
\end{equation}
Equations~\eqref{eq:supp_multi_time_factorization} and \eqref{eq:supp_multi_time_relative_part} establish the Wegner-flow normal form of the $n$-th unequal-time correlation.

\subsection{Two-time correlations}
The two-time result follows by setting $M=2$.
Let the future time slice contain $p$ legs with wave vectors $\vec k_1,\cdots,\vec k_p$, and the past time slice contain $q$ legs with wave vectors $\vec q_1,\cdots,\vec q_q$.
We define
\begin{equation}
	\vec K
	=
	\sum_{i=1}^{p}\vec k_i
	=
	-
	\sum_{j=1}^{q}\vec q_j .
	\label{eq:supp_two_time_K_def}
\end{equation}
Then Eq.~\eqref{eq:supp_multi_time_factorization} reduces to
\begin{equation}
	C_{p,q}^{\mathrm E}(\tau)
	=
	\ee^{-\ii\vec U\cdot\vec K\tau}
	\ee^{-\frac12 V_0^2K^2\tau^2}
	\ee^{-D_0 K^2 \tau}
	C_{p,q}^{\mathrm{rel}}(\tau),
	\label{eq:supp_two_time_factorization}
\end{equation}
where the relative part satisfies
\begin{equation}
	C_{p,q}^{\mathrm{rel}}(X,\tau;Y,0)
	=
	\ee^{\tau\mathcal M_X}
	C_{p+q}(X,Y)
	+
	\int_0^{\tau}
	\dif s\,
	\ee^{(\tau-s)\mathcal M_X}
	\varPhi_p^X
	C_{p-2,q}^{\mathrm{rel}}(X,s;Y,0).
	\label{eq:supp_two_time_relative_duhamel}
\end{equation}
Here, $\mathcal M_X$ is the operator defined in Eq.~\eqref{eq:supp_active_M_def} with active set $A=X$, and $\varPhi_p^X$ is the forcing source acting only on the active future coordinates $X$.
This source is the unequal-time analogue of the equal-time source $\varPhi_n$ in Eq.~\eqref{eq:supp_force_hierarchy_H}.
Explicitly, it satisfies
\begin{equation}
	\varPhi_p^X
	C_{p-2,q}^{\mathrm{rel}}(X,s;Y,0)
	=
	\sum_{1\le a<b\le p}
	\chi(\vec x_a-\vec x_b)
	C_{p-2,q}^{\mathrm{rel}}
	(X^{\hat a\hat b},s;Y,0),
	\label{eq:supp_two_time_relative_source}
\end{equation}
where $X^{\hat a\hat b}$ denotes the set obtained by removing the two coordinates $\vec x_a$ and $\vec x_b$ from $X$.

The corresponding zero-mode contribution is
\begin{equation}
	C_{p,q}^{\mathrm{rel},\mathrm{zm}}(X,\tau;Y,0)
	=
	\ee^{\tau\mathcal M_X}
	Z_{p+q}(X,Y).
	\label{eq:supp_two_time_relative_zm}
\end{equation}
Combining Eqs.~\eqref{eq:supp_two_time_factorization} and \eqref{eq:supp_two_time_relative_zm} yields the Eulerian zero-mode contribution
\begin{equation}
	C_{p,q}^{\mathrm E,\mathrm{zm}}(\tau)
	=
	\ee^{-\ii\vec U\cdot\vec K\tau}
	\ee^{-\frac12 V_0^2K^2\tau^2}
	\ee^{-D_0 K^2 \tau}
	\left[
	\ee^{\tau\mathcal M_X}
	Z_{p+q}
	\right],
	\label{eq:supp_two_time_E_zm_final}
\end{equation}
where the square bracket indicates that the relative propagator acts on the future coordinates $X$ of the equal-time zero mode.

Equation~\eqref{eq:supp_two_time_E_zm_final} factorizes the unequal-time zero mode into a kinematic Eulerian part and an intrinsic relative part.
The first three factors depend only on the total wave vector $\vec K$ in the future time slice. 
The mean flow produces a phase, the frozen sweeping velocity produces Gaussian decorrelation, and the zero-separation covariance $D_{ij}(\vec 0)$ produces center diffusion.
The bracketed term contains the relative dynamics.
It follows from taking the equal-time zero mode $Z_{p+q}(X,Y)$ as the boundary value at $\tau=0^+$ and evolving only the future coordinates $X$ with the active-set generator $\mathcal M_X$.
Thus, the unequal-time zero mode introduces no new spatial exponent. 
Its anomalous scaling is inherited from $Z_{p+q}$, whereas its temporal dependence is generated by the semigroup $\ee^{\tau\mathcal M_X}$.
For a generic Eulerian observable with $\vec K\ne\vec 0$, sweeping and center diffusion obscure this intrinsic dynamics.
It is exposed either by projection onto the sector $\vec K=\vec 0$ or by removal of the Eulerian sweeping contribution, for example in a quasi-Lagrangian formulation~\cite{Belinicher1987_A}.

We then validate the general unequal-time normal form in Eq.~\eqref{eq:supp_multi_time_factorization} by recovering the elementary two-point limit.
Because this limit contains only a single active scalar factor at the future time, it isolates the one-particle sector of the Wegner normal form.
By setting $p=q=1$, we define the two-time spectral correlation as
\begin{equation}
	C_{1,1}^{\mathrm E}(\vec k,\tau)
	=
	\langle
	\hat\theta_{\vec k}(t+\tau)
	\hat\theta_{-\vec k}(t)
	\rangle,
	\quad
	\tau>0 .
	\label{eq:supp_C11_E_def}
\end{equation}
The future time slice carries the total wave vector $\vec K=\vec k$.
Equation~\eqref{eq:supp_two_time_factorization} then reduces to
\begin{equation}
	C_{1,1}^{\mathrm E}(\vec k,\tau)
	=
	\ee^{-\ii\vec U\cdot\vec k\tau}
	\ee^{-\frac12 V_0^2k^2\tau^2}
	\ee^{-D_0 k^2 \tau}
	C_{1,1}^{\mathrm{rel}}(\vec k,\tau).
	\label{eq:supp_C11_factorization}
\end{equation}
Since only a single active scalar factor exists at the future time, no scalar-forcing pair forms within the future time slice.
Consequently, the source term vanishes, rendering the relative part purely homogeneous, namely
\begin{equation}
	C_{1,1}^{\mathrm{rel}}(\vec x,\tau;\vec y,0)
	=
	\ee^{\tau\mathcal M_{\vec x}}
	C_2(\vec x,\vec y).
	\label{eq:supp_C11_relative_space}
\end{equation}
Since $d_{ij}(\vec 0)=0$, the relative generator for a single active coordinate reduces to $\mathcal M_{\vec x}=\varkappa\nabla_{\vec x}^{2}$.
In Fourier space, this yields
\begin{equation}
	C_{1,1}^{\mathrm{rel}}(\vec k,\tau)
	=
	\ee^{-\varkappa k^2\tau}
	C_2(\vec k).
	\label{eq:supp_C11_relative_fourier}
\end{equation}
Combining Eqs.~\eqref{eq:supp_C11_factorization} and \eqref{eq:supp_C11_relative_fourier}, we obtain
\begin{equation}
	C_{1,1}^{\mathrm E}(\vec k,\tau)
	=
	\ee^{-\ii\vec U\cdot\vec k\tau}
	\ee^{-\frac12 V_0^2k^2\tau^2}
	\ee^{-(\varkappa + D_0) k^2 \tau}
	C_2(\vec k).
	\label{eq:supp_C11_final}
\end{equation}
In the limit $\boldsymbol{U}=\boldsymbol{V}=0$, Eq.~\eqref{eq:supp_C11_final} reduces to the standard Kraichnan-model result~\cite{Yang2023_Space}, and its inviscid limit follows upon taking $\varkappa\to0$, confirming the self-consistency of the present formalism.
This result serves as a two-point validation of the unequal-time normal form.
Specifically, the mean flow contributes a phase factor, the frozen sweeping velocity leads to a Gaussian decorrelation, and the diagonal one-particle Wegner self-energy determines the diffusive decay rate $(\varkappa + D_0) k^2$.
At this order, no many-body dynamic zero mode appears.

\subsection{Second-order space-time correlation}
We then obtain the normalized second-order space-time correlation function
\begin{equation}
    R(r,\tau)
    :=
    \frac{C^{\mathrm{E}}_{1,1}(r,\tau)}{C_2(0)}.
\end{equation}
Henceforth, we set $d=3$ and, without loss of generality, choose $\vec{r}=(r,0,0)$ and $\vec{U}=(U,0,0)$.
We define the one-particle spreading length
\begin{equation}
    \ell_{\mathrm{K}}^2(\tau)
    =
    V_0^2\tau^2+2(\varkappa+D_0)\tau
    \label{eq:ell_K}
\end{equation}
through the displacement variance in each Cartesian direction, which denotes the per-component variance rather than the total three-dimensional mean-square displacement.
Equation~\eqref{eq:supp_C11_final} then simplifies to
\begin{equation}
    C^{\mathrm{E}}_{1,1}(\boldsymbol{k},\tau)
    =
    \mathrm{e}^{-\mathrm{i}Uk_1\tau}
    \mathrm{e}^{-\ell_{\mathrm{K}}^2(\tau)k^2/2}
    C_2(\boldsymbol{k}).
    \label{eq:C11_streamwise_spectral}
\end{equation}

Taking the inverse Fourier transform of Eq.~\eqref{eq:C11_streamwise_spectral} yields
\begin{equation}
    R(r,\tau)
    =
    \int_{\mathbb{R}^3}
    \frac{\mathrm{d}^3\boldsymbol{\rho}}
    {[2\pi\ell_{\mathrm{K}}^2(\tau)]^{3/2}}
    \exp\bigg(
        -\frac{\rho^2}{2\ell_{\mathrm{K}}^2(\tau)}
    \bigg)
    R
    \left(
        \left|
        (r-U\tau)\hat{\boldsymbol{e}}
        -\boldsymbol{\rho}
        \right|,
        0
    \right),
    \label{eq:C11_gaussian_convolution}
\end{equation}
where $\hat{\boldsymbol{e}}$ is the streamwise unit vector.
In the inertial-convective range, the equal-time correlation takes the Kraichnan form
\begin{equation}
    R(r,0)
    =
    1-\left(\frac{r}{L_\xi}\right)^{2-\xi},
    \label{eq:C2_inertial}
\end{equation}
with the scalar correlation scale given by
\begin{equation}
    L_\xi
    =
    \left[
        \frac{12D_1(2-\xi)C_2(0)}
        {F_0}
    \right]^{1/(2-\xi)}.
    \label{eq:Lxi}
\end{equation} 
Substituting Eq.~\eqref{eq:C2_inertial} into Eq.~\eqref{eq:C11_gaussian_convolution} and evaluating the Gaussian moment yield
\begin{equation}
    R(r,\tau)
    = 1-
    \bigg(
        \frac{C_\xi\ell_{\mathrm{K}}(\tau)}
        {L_\xi}
    \bigg)^{2-\xi}
    {}_1F_1\bigg(
        \frac{\xi}{2}-1;
        \frac{3}{2};
        -\frac{(r-U\tau)^2}
        {2\ell_{\mathrm{K}}^2(\tau)}
    \bigg),
    \label{eq:C11_explicit}
\end{equation}
with a constant
\begin{equation}
    C_\xi
    =
    \sqrt{2}
    \bigg[
        \frac{2}{\sqrt{\pi}}
        \Gamma\bigg(\frac{5-\xi}{2}\bigg)
    \bigg]^{1/(2-\xi)}.
    \label{eq:Cxi}
\end{equation}
Under the Kolmogorov-Obukhov-Corrsin scaling law of scalar fluctuations~\cite{Corrsin1951_on}, i.e., $\xi=4/3$, this constant evaluates to $C_\xi\approx 1.55$.
Equation~\eqref{eq:C11_explicit} reproduces the exact result of Ref.~\cite{Wang2025_Solvable} and provides a two-point check of the unequal-time normal form.

\subsection{Minimal non-Gaussian dynamic zero mode}\label{sec:supp_minimal_dynamic_zero_mode}
At fourth order, the first genuinely many-body unequal-time zero mode emerges.
The minimal non-Gaussian observable comprises two scalar factors at the future time and two at the past time.
Let $X=(\vec x_1,\vec x_2)$, $Y=(\vec y_1,\vec y_2)$, and $\tau>0$.
For $p=q=2$, the relative zero-mode contribution obtained from Eq.~\eqref{eq:supp_two_time_relative_zm} is
\begin{equation}
	C_{2,2}^{\mathrm{rel},\mathrm{zm}}(X,\tau;Y,0)
	=
	\ee^{\tau\mathcal M_X}
	Z_4(X,Y).
	\label{eq:supp_C22_relative_zm}
\end{equation}
Here, $Z_4$ denotes the leading equal-time fourth-order zero mode, and $\mathcal M_X$ acts only on the two future coordinates in $X$.
Consequently, the anomalous spatial scaling is governed by the equal-time zero mode, whereas the temporal dependence is generated by the two-particle relative propagator.

Equation~\eqref{eq:supp_C22_relative_zm} also admits a convenient frequency-domain representation.
We define the one-sided transform of the relative zero-mode contribution as
\begin{equation}
	\widetilde C_{2,2}^{\mathrm{rel},\mathrm{zm}}(X,Y;\omega)
	=
	\int_0^\infty
	\dif\tau\,
	\ee^{\ii\omega\tau}
	C_{2,2}^{\mathrm{rel},\mathrm{zm}}(X,\tau;Y,0).
	\label{eq:supp_C22_transform_def}
\end{equation}
Applying Eq.~\eqref{eq:supp_C22_relative_zm}, we express this transform as
\begin{equation}
	\widetilde C_{2,2}^{\mathrm{rel},\mathrm{zm}}(X,Y;\omega)
	=
	R_X(\omega)Z_4(X,Y),
	\label{eq:supp_C22_resolvent}
\end{equation}
with the active two-particle resolvent
\begin{equation}
	R_X(\omega)
	=
	\left(
	0^+
	-
	\ii\omega
	-
	\mathcal M_X
	\right)^{-1}.
	\label{eq:supp_RX_def}
\end{equation}
Here, $\omega$ is conjugate to the time separation $\tau$, and $0^+$ enforces the retarded prescription.
Equation~\eqref{eq:supp_C22_resolvent} is the frequency-domain representation of the semigroup $\ee^{\tau\mathcal M_X}$, which demonstrates that the unequal-time spectrum is obtained by probing the equal-time four-particle zero mode with the response of the active two-particle generator.
Consequently, the spatial anomalous exponent is inherited from $Z_4$, whereas the frequency dependence is governed by the spectrum of $\mathcal M_X$.
This formulation constitutes the simplest dynamic zero-mode normal form.

Then, we project the four-point correlation function onto a two-time square-increment observable.
For a given separation vector $\vec r$, we define the scalar increment $\delta_{\vec r}\theta(\vec x,t)=\theta(\vec x+\vec r,t)-\theta(\vec x,t)$, and the corresponding two-time square-increment correlation
\begin{equation}
	F_{2,2}(\vec r,\tau)
	=
	\left\langle
	[\delta_{\vec r}\theta(\vec x,t+\tau)]^2
	[\delta_{\vec r}\theta(\vec x,t)]^2
	\right\rangle .
	\label{eq:supp_F22_vec_def}
\end{equation}
In the isotropic sector considered here, the ensemble-averaged correlation is invariant under simultaneous rotations of the separation vector, namely $F_{2,2}(\mathcal R\vec r,\tau)=F_{2,2}(\vec r,\tau)$ for any rotation $\mathcal R\in \mathrm{SO}(d)$.
We therefore choose an arbitrary fixed unit vector $\hat{\vec e}$ and define
\begin{equation}
	F_{2,2}(r,\tau)
	:=
	F_{2,2}(r\hat{\vec e},\tau),
	\quad
	r>0 .
	\label{eq:supp_F22_r_def}
\end{equation}

Let $\mathcal P_r$ denote the square-increment projection of a two-time four-point function associated with Eq.~\eqref{eq:supp_F22_r_def} (detailed in Sec.~\ref{app:supp_increment_projection}).
The relative zero-mode contribution is then given by
\begin{equation}\label{eq:supp_F22_projection}
	F_{2,2}^{\mathrm{rel},\mathrm{zm}}(r,\tau)
	=
	\mathcal P_r
	\left[
	\ee^{\tau\mathcal M_X}Z_4
	\right].
\end{equation}
The operator $\mathcal P_r$ fixes the separation of both the future and past pairs to the magnitude $r$, extracting the signed combination generated by the two scalar increments.

We now derive an explicit expression for $F_{2,2}^{\mathrm{rel},\mathrm{zm}}$ within the inertial range.
To this end, we first establish the radial form of the active two-particle generator.
For the active set $X=(\vec x_1,\vec x_2)$, neglecting molecular diffusion in the inertial range simplifies Eq.~\eqref{eq:supp_active_M_def} to
\begin{equation}
	\mathcal M_X
	= -\sum_{a,b=1}^{2}
	d_{ij}(\vec x_a-\vec x_b)
	\partial_{x_a^i}\partial_{x_b^j}.
	\label{eq:app_MX_two_particle}
\end{equation}
We introduce the relative coordinate $\vec r = \vec x_1-\vec x_2$ with $r=|\vec r|$ and $\hat r_i=r_i/r$.
For a radial function $f(r)$, the derivatives with respect to the two active coordinates satisfy $\partial_{x_1^i}f = \partial_{r_i}f$ and $\partial_{x_2^i}f = -\partial_{r_i}f$.
Substituting $d_{ij}(\vec 0)=0$ and the symmetry relation $d_{ij}(-\vec r) = d_{ij}(\vec r)$ for a statistically isotropic velocity field into Eq.~\eqref{eq:app_MX_two_particle} yields
\begin{equation}
	\mathcal M_X f
	=
	2d_{ij}(\vec r)
	\partial_{r_i}\partial_{r_j}f.
	\label{eq:app_MX_relative_tensor}
\end{equation}
Then, substituting Eq.~\eqref{eq:supp_dij_def_H} and the Hessian
\begin{equation}
	\partial_{r_i}\partial_{r_j}f(r)
	=
	\left(
	\delta_{ij}-\hat r_i\hat r_j
	\right)
	\frac{f'(r)}{r}
	+
	\hat r_i\hat r_j f''(r)
	\label{eq:app_radial_hessian}
\end{equation}
into Eq.~\eqref{eq:app_MX_relative_tensor} yields the active two-particle generator
\begin{equation}\label{eq:supp_MX_radial}
	\mathcal M_X f
	=
	2D_1(d-1)r^\xi
	\left[
	f''(r)
	+
	(d - 1 +\xi)\frac{f'(r)}{r}
	\right].
\end{equation}

The inertial-range scale covariance of the active generator, as proven in Sec.\ref{app:supp_dynamic_scaling_form}, implies the scaling form
\begin{equation}\label{eq:supp_F22_ansatz}
	F_{2,2}^{\mathrm{rel},\mathrm{zm}}(r,\tau)
	=
	A_4 r^{\zeta_4}\varPhi_{2,2}(s),
	\quad
	s=
	\frac{D_1\tau}{r^z},
	\quad
	z=2-\xi,
\end{equation}
with a nonuniversal amplitude $A_4$, the equal-time fourth-order anomalous exponent $\zeta_4$, and the normalization constraint $\varPhi_{2,2}(0)=1$.
Substituting Eq.~\eqref{eq:supp_F22_ansatz} into the isotropic radial closure
\begin{equation}
	\partial_\tau F_{2,2}^{\mathrm{rel},\mathrm{zm}}
	=
	\mathcal M_X F_{2,2}^{\mathrm{rel},\mathrm{zm}}
	\label{eq:supp_F22_dynamic_equation}
\end{equation}
yields
\begin{equation}
	\varPhi_{2,2}'(s)
	=
	2(d-1)
	\left[
	z^2s^2\varPhi_{2,2}''(s)
	-
	z(2\zeta_4+d-2z)s\varPhi_{2,2}'(s)
	+
	\zeta_4(\zeta_4+d-z)\varPhi_{2,2}(s)
	\right].
	\label{eq:supp_Phi22_ODE}
\end{equation}
Solving the ordinary differential equation in Eq.~\eqref{eq:supp_Phi22_ODE} as detailed in Sec.~\ref{app:supp_exact_Phi22_solution} yields the analytical solution
\begin{equation}\label{S-eq:varPhi22_sol_exact}
	\varPhi_{2,2}(s)
	=
	\frac{
	\Gamma\!\left(\dfrac{d+\zeta_4}{z}\right)
	}{
	\Gamma\!\left(\dfrac{d}{z}\right)
	}
	\bigg(\frac{1}{2(d-1)z^2s} \bigg)^{-\zeta_4/z}
	{}_1F_1
	\left(
	-\frac{\zeta_4}{z};
	\frac{d}{z};
	-\frac{1}{2(d-1)z^2s}
	\right).
\end{equation}
This result constitutes the minimal explicit unequal-time prediction of the dynamic zero-mode theory, thereby providing a zero-mode interpretation of dynamic scaling in passive-scalar turbulence~\cite{Mitra2005_Dynamics}.
The coefficient of the first-order temporal correction is determined by the two-particle relative generator and the equal-time anomalous exponent.
We emphasize that the exponent $\zeta_4$ is the homogeneous exponent of the leading equal-time fourth-order zero mode and is independent of the unequal-time propagation.

\section{Appendix: Mathematical details}

\subsection{Proof of Eq.~\eqref{eq:supp_telescoping_identity}}\label{sec:proof_supp_telescoping_identity}
We first introduce the total wave vector
\begin{equation}
	\vec P_i
	=
	\sum_{j=1}^{n_i}
	\vec k_{i,j},
	\quad
	i=1,\cdots,M
	\label{eq:supp_time_slice_momentum}
\end{equation}
carried by the $i$-th time slice.
With this notation, Eq.~\eqref{eq:supp_active_Km_def} becomes
\begin{equation}
	\vec K_m
	=
	\sum_{i=m}^{M}
	\vec P_i,
	\quad
	m=2,\cdots,M.
	\label{eq:supp_Km_Pi}
\end{equation}
Subsequently, we obtain
\begin{equation}\label{eq:supp_telescoping_step_1}
    \begin{aligned}
        \sum_{m=2}^{M}
    	\vec K_m\Delta t_m
    	&=
    	\sum_{m=2}^{M}
    	\left(
    	\sum_{i=m}^{M}
    	\vec P_i
    	\right)
    	(t_m-t_{m-1})
        \\
        &=
    	\sum_{i=2}^{M}
    	\vec P_i
    	\sum_{m=2}^{i}
    	(t_m-t_{m-1})
        \\
        &= \sum_{i=2}^{M}
	    \vec P_i(t_i-t_1)
        \\
        &=\sum_{i=2}^{M}
    	\vec P_i t_i
    	-
    	t_1
    	\sum_{i=2}^{M}
    	\vec P_i.
    \end{aligned}
\end{equation}
For homogeneous correlations, Eq.~\eqref{eq:supp_multi_time_total_momentum} imposes a vanishing total wave vector, $\sum_{i=1}^{M} \vec P_i =\vec 0$.
Consequently, we have
\begin{equation}
	\sum_{i=2}^{M}
	\vec P_i
	=
	-\vec P_1 .
	\label{eq:supp_Pi_momentum_balance}
\end{equation}
Substituting Eq.~\eqref{eq:supp_Pi_momentum_balance} into Eq.~\eqref{eq:supp_telescoping_step_1} yields
\begin{equation}
	\sum_{m=2}^{M}
	\vec K_m\Delta t_m
	=
	\sum_{i=2}^{M}
	\vec P_i t_i
	+
	\vec P_1 t_1
	=
	\sum_{i=1}^{M}
	\vec P_i t_i .
	\label{eq:supp_telescoping_step_5}
\end{equation}
Finally, using the definition of $\vec P_i$ in Eq.~\eqref{eq:supp_time_slice_momentum}, we obtain Eq.~\eqref{eq:supp_telescoping_identity}.
This identity shows that the accumulated sweeping phase depends only on the external wave vectors and their observation times.
It is independent of the arbitrary choice of the earliest reference time because the total wave vector vanishes.



\subsection{Definition of Eq.~\eqref{eq:supp_F22_projection}}\label{app:supp_increment_projection}
We define the projection from the two-time, four-point correlation function to the square-increment observable.
For a fixed separation vector $\vec r$, we define the spatial coordinates $\vec x_0 = \vec x$ and $\vec x_1 = \vec x+\vec r$, together with the sign factors $s_0=-1$ and $s_1=1$.
The scalar increment is then expressed as
\begin{equation}
	\delta_{\vec r}\theta(\vec x,t)
	=
	\sum_{\sigma=0}^{1}
	s_\sigma
	\theta(\vec x_\sigma,t)
	=
	\theta(\vec x+\vec r,t)-\theta(\vec x,t).
	\label{eq:app_increment_signed_sum}
\end{equation}

We consider an arbitrary two-time, four-point function $G(X,\tau;Y,0) = G(\vec x_1,\vec x_2,\tau; \vec y_1,\vec y_2,0)$, where $X=(\vec x_1,\vec x_2)$ and $Y=(\vec y_1,\vec y_2)$.
Expanding the product of the two squared increments yields
\begin{equation}
	[\delta_{\vec r}\theta(\vec x,t+\tau)]^2
	[\delta_{\vec r}\theta(\vec x,t)]^2
	=
	\sum_{\sigma_1,\sigma_2,\sigma_3,\sigma_4=0}^{1}
	s_{\sigma_1}s_{\sigma_2}s_{\sigma_3}s_{\sigma_4}
	\theta(\vec x_{\sigma_1},t+\tau)
	\theta(\vec x_{\sigma_2},t+\tau)
	\theta(\vec x_{\sigma_3},t)
	\theta(\vec x_{\sigma_4},t).
	\label{eq:app_increment_product_expansion}
\end{equation}
This identity motivates the definition of the square-increment projection
\begin{equation}
	\mathcal P_{\vec r}G
	=
	\sum_{\sigma_1,\sigma_2,\sigma_3,\sigma_4=0}^{1}
	s_{\sigma_1}s_{\sigma_2}s_{\sigma_3}s_{\sigma_4}
	G(
	\vec x_{\sigma_1},\vec x_{\sigma_2},\tau;
	\vec x_{\sigma_3},\vec x_{\sigma_4},0
	).
	\label{eq:app_Pr_vec_def}
\end{equation}
The summation in Eq.~\eqref{eq:app_Pr_vec_def} comprises sixteen terms generated by the two future and two past increments.
The corresponding sign factors cancel contributions that are independent of either endpoint of the separation vector.

For statistically homogeneous correlations, the correlation function is independent of the reference position $\vec x$.
In the statistically isotropic sector considered here, this projection satisfies $\mathcal P_{\mathcal R\vec r}G = \mathcal P_{\vec r}G$ or any $\mathcal R\in\mathrm{SO}(d)$.
Consequently, the projected correlation depends only on the magnitude $r=|\vec r|$.
We therefore write $\mathcal P_r G := \mathcal P_{\vec r}G$, where statistical isotropy ensures that the right-hand side is independent of the direction of $\vec r$.
Applying this operator to the relative dynamic zero mode yields
\begin{equation}
	F_{2,2}^{\mathrm{rel},\mathrm{zm}}(r,\tau)
	=
	\mathcal P_r
	C_{2,2}^{\mathrm{rel},\mathrm{zm}}(X,\tau;Y,0).
	\label{eq:app_F22_projection_general}
\end{equation}
Substituting Eq.~\eqref{eq:supp_C22_relative_zm} into Eq.~\eqref{eq:app_F22_projection_general}, we obtain Eq.~\eqref{eq:supp_F22_projection}.
The operator $\mathcal M_X$ acts exclusively on the future coordinates $X$.
Conversely, the past coordinates $Y$ remain fixed parameters during the propagation from $t$ to $t+\tau$.

Next, we express the semigroup $\{\ee^{\tau \mathcal{M}_X}\}_{\tau\ge 0}$ in terms of its integral kernel.
Let $\Omega$ denote the spatial domain.
For an unbounded system, $\Omega=\mathbb R^d$, whereas for a periodic system, $\Omega$ is the $d$-dimensional torus associated with the periodic box.
We define the Green function $G_X(X,X';\tau)$ by
\begin{equation}
	\left[
	\mathrm{e}^{\tau\mathcal M_X}f
	\right](X)
	=
	\int_{\Omega^2}
	\dif X'\,
	G_X(X,X';\tau)
	f(X'),
	\quad
	\tau>0,
	\label{eq:app_GX_kernel_def}
\end{equation}
where $X'=(\vec x_1',\vec x_2')$ is the configuration and $\int_{\Omega^2}\dif X'=\int_{\Omega}\dif \vec{x}_1'\int_{\Omega}\dif \vec{x}_2'$ is the corresponding integration measure.
Consequently, the integration extends over all possible initial configurations of the active future pair.

The kernel $G_X$ satisfies the forward equation
\begin{equation}\label{eq:app_GX_forward_equation}
    \begin{cases}
        (
    	\partial_\tau
    	-
    	\mathcal M_X^{(X)}
    	)
    	G_X(X,X';\tau)
    	=
    	0,
    	\quad
    	\tau>0, \\
        G_X(X,X';0^+)
        =
        \delta_{\Omega}(X-X').
    \end{cases}
\end{equation}
Here, the superscript ``$(X)$'' indicates that $\mathcal M_X$ acts on the unprimed coordinates, and $\delta_{\Omega}^{(d)}$ denotes the Dirac delta function on the domain $\Omega$.
The retarded kernel is defined to vanish for $\tau<0$.

The Green function satisfies the semigroup property
\begin{equation}
	G_X(X,X'';\tau_1+\tau_2)
	=
	\int_{\Omega^2}
	\dif X'\,
	G_X(X,X';\tau_1)
	G_X(X',X'';\tau_2),
	\quad
	\tau_1,\tau_2>0.
	\label{eq:app_GX_semigroup_property}
\end{equation}
If the boundary conditions preserve probability and $\mathcal M_X1=0$, the kernel also satisfies the normalization relation
\begin{equation}
	\int_{\Omega^2}
	\dif X'\,
	G_X(X,X';\tau)
	=
	1.
	\label{eq:app_GX_normalization}
\end{equation}
In the Markovian representation of the white-in-time model, $G_X(X,X';\tau)$ serves as the transition kernel that propagates the active pair from the configuration $X'$ at the fusion time to the configuration $X$ after a time interval $\tau$.

Using Eq.~\eqref{eq:app_GX_kernel_def}, the dynamic four-point zero mode is expressed as
\begin{equation}
	C_{2,2}^{\mathrm{rel},\mathrm{zm}}
	(X,\tau;Y,0)
	=
	\int_{\Omega^2}
	\dif X'\,
	G_X(X,X';\tau)
	Z_4(X',Y).
	\label{eq:app_Green_C22}
\end{equation}
Here, the coordinate $X'$ labels the future pair at the equal-time boundary $\tau=0^+$.
The coordinate $X$ labels the same active pair at the subsequent observation time.
The coordinate $Y$ labels the past pair and remain unintegrated because they are no longer active.
Applying the square-increment projection to Eq.~\eqref{eq:app_Green_C22} yields
\begin{equation}
	F_{2,2}^{\mathrm{rel},\mathrm{zm}}(r,\tau)
	=
	\sum_{\sigma_1,\sigma_2,\sigma_3,\sigma_4=0}^{1}
	s_{\sigma_1}s_{\sigma_2}s_{\sigma_3}s_{\sigma_4}
	\int_{\Omega^2}
	\dif X'\,
	G_X(
	X_{\sigma_1\sigma_2},
	X';
	\tau
	)
	Z_4(
	X',
	Y_{\sigma_3\sigma_4}
	),
	\label{eq:app_Green_F22_explicit}
\end{equation}
where $X_{\sigma_1\sigma_2}=(\vec x_{\sigma_1},\vec x_{\sigma_2})$ and $Y_{\sigma_3\sigma_4}=(\vec x_{\sigma_3},\vec x_{\sigma_4})$.
Equivalently, we obtain
\begin{equation}
	F_{2,2}^{\mathrm{rel},\mathrm{zm}}(r,\tau)
	=
	\mathcal P_r
	\int_{\Omega^2}
	\dif X'\,
	G_X(X,X';\tau)
	Z_4(X',Y).
	\label{eq:app_Green_F22}
\end{equation}
Equation~\eqref{eq:app_Green_F22} separates the two key components of the dynamic zero mode.
Specifically, the equal-time four-particle zero mode $Z_4$ provides the anomalous spatial boundary data.
Meanwhile, the two-particle Green function governs the subsequent relative evolution of the active future pair.
The square-increment projection then extracts the experimentally accessible fourth-order observable from the propagated four-point function.

In the equal-time limit, Eq.~\eqref{eq:app_Green_F22} reduces to
\begin{equation}
	\lim_{\tau\to0^+}
	F_{2,2}^{\mathrm{rel},\mathrm{zm}}(r,\tau)
	=
	\mathcal P_r
	\int_{\Omega^2}
	\dif X'\,
	\delta_{\Omega}(X-X')
	Z_4(X',Y)
	=
	\mathcal P_rZ_4(X,Y).
	\label{eq:app_F22_equal_time_limit}
\end{equation}
Consequently, the unequal-time square-increment zero mode continuously reduces to its equal-time fourth-order counterpart.

\subsection{Proof of Eq.~\eqref{eq:supp_F22_ansatz}}\label{app:supp_dynamic_scaling_form}
Using the scale covariance of the relative generator, we derive the inertial-range scaling form of the fourth-order dynamic zero mode in Eq.~\eqref{eq:supp_F22_ansatz}.
This derivation applies to the zero-mode sector at separations well above the diffusive scale and well below the forcing scale.
Consequently, the molecular diffusion term is omitted.

We consider the fourth-order relative dynamic zero mode in Eq.~\eqref{eq:supp_C22_relative_zm}.
In the inertial range, the velocity structure tensor is homogeneous, satisfying
\begin{equation}
	d_{ij}(\lambda\vec r)
	=
	\lambda^\xi d_{ij}(\vec r),
	\quad
	\lambda>0 .
	\label{eq:app_dij_homogeneity}
\end{equation}
We define the dilation operator as
\begin{equation}
	\left(
	\mathcal S_\lambda f
	\right)(X,Y)
	:=
	f(\lambda X,\lambda Y),
	\label{eq:app_dilation_operator}
\end{equation}
where $\lambda X = (\lambda\vec x_1,\lambda\vec x_2)$ and $\lambda Y = (\lambda\vec y_1,\lambda\vec y_2)$.
Under this dilation, the spatial derivatives transform as $\partial_{x_a^i}(\mathcal S_\lambda f ) = \lambda \mathcal S_\lambda( \partial_{x_a^i}f )$ and $\partial_{x_a^i}\partial_{x_b^j} (\mathcal S_\lambda f ) = \lambda^2 \mathcal S_\lambda ( \partial_{x_a^i}\partial_{x_b^j}f )$.
Substituting Eq.~\eqref{eq:app_dij_homogeneity} into Eq.~\eqref{eq:app_MX_two_particle}, we obtain
\begin{equation}
	\begin{aligned}
	\mathcal M_X\mathcal S_\lambda f
	&=
	-
	\sum_{a,b=1}^{2}
	d_{ij}(\vec x_a-\vec x_b)
	\partial_{x_a^i}\partial_{x_b^j}
	\left(
	\mathcal S_\lambda f
	\right)
	\\
	&=
	-\lambda^{2-\xi}
	\mathcal S_\lambda
	\bigg[
	\sum_{a,b=1}^{2}
	d_{ij}(\vec x_a-\vec x_b)
	\partial_{x_a^i}\partial_{x_b^j}f
	\bigg],
	\end{aligned}
	\label{eq:app_M_scaling_steps}
\end{equation}
which yields the scale covariance relation
\begin{equation}
	\mathcal M_X\mathcal S_\lambda
	=
	\lambda^z
	\mathcal S_\lambda\mathcal M_X,
	\quad
	z=2-\xi .
	\label{eq:app_M_scale_covariance}
\end{equation}
The exponent $z$ is thus determined by the scaling dimension of the inertial-range generator.

Substituting Eq.~\eqref{eq:app_M_scale_covariance} into the power-series expansion of the exponential $\ee^{\tau\mathcal M_X} = \sum_{n=0}^{\infty} \frac{\tau^n}{n!} \mathcal M_X^n$ yields
\begin{equation}
	\mathcal S_\lambda
	\ee^{\tau\mathcal M_X}
	=
	\ee^{\tau\lambda^{-z}\mathcal M_X}
	\mathcal S_\lambda.
	\label{eq:app_semigroup_scale_covariance}
\end{equation}
Besides, the leading equal-time fourth-order zero mode is homogeneous with exponent $\zeta_4$, satisfying
\begin{equation}
	\mathcal S_\lambda Z_4
	=
	Z_4(\lambda X,\lambda Y)
	=
	\lambda^{\zeta_4}
	Z_4(X,Y).
	\label{eq:app_Z4_scale_homogeneity}
\end{equation}
Applying $\mathcal S_\lambda$ to Eq.~\eqref{eq:supp_C22_relative_zm} and invoking Eqs.~\eqref{eq:app_semigroup_scale_covariance} and \eqref{eq:app_Z4_scale_homogeneity}, we obtain
\begin{equation}
	\begin{aligned}
	C_{2,2}^{\mathrm{rel},\mathrm{zm}}
	(\lambda X,\tau;\lambda Y,0)
	&=
	\mathcal S_\lambda
	\left[
	\ee^{\tau\mathcal M_X}Z_4
	\right]
	\\
	&=
	\ee^{\tau\lambda^{-z}\mathcal M_X}
	\mathcal S_\lambda Z_4
	\\
	&=
	\lambda^{\zeta_4}
	C_{2,2}^{\mathrm{rel},\mathrm{zm}}
	\left(
	X,\lambda^{-z}\tau;Y,0
	\right).
	\end{aligned}
	\label{eq:app_C22_scaling_steps}
\end{equation}
Equation \eqref{eq:app_C22_scaling_steps} is equivalently expressed as
\begin{equation}
	C_{2,2}^{\mathrm{rel},\mathrm{zm}}
	(\lambda X,\lambda^z\tau;\lambda Y,0)
	=
	\lambda^{\zeta_4}
	C_{2,2}^{\mathrm{rel},\mathrm{zm}}
	(X,\tau;Y,0).
	\label{eq:app_C22_dynamic_homogeneity}
\end{equation}

We next project Eq.~\eqref{eq:app_C22_dynamic_homogeneity} onto the isotropic two-time square-increment observable.
Letting $\mathcal P_r$ denote the signed four-point projection at future and past pair separation $r$, the projected zero-mode contribution is given by
\begin{equation}
	F_{2,2}^{\mathrm{rel},\mathrm{zm}}(r,\tau)
	=
	\mathcal P_r
	C_{2,2}^{\mathrm{rel},\mathrm{zm}}(X,\tau;Y,0).
	\label{eq:app_F22_projection_def}
\end{equation}
Under a uniform dilation of all coordinates, the pair separation scales from $r$ to $\lambda r$, meaning the projection operator satisfies
\begin{equation}
	\mathcal P_{\lambda r}
	=
	\mathcal P_r\mathcal S_\lambda .
	\label{eq:app_projection_scale_covariance}
\end{equation}
Combining Eqs.~\eqref{eq:app_C22_dynamic_homogeneity} and \eqref{eq:app_projection_scale_covariance} yields
\begin{equation}
	\begin{aligned}
	F_{2,2}^{\mathrm{rel},\mathrm{zm}}
	(\lambda r,\lambda^z\tau)
	&=
	\mathcal P_{\lambda r}
	C_{2,2}^{\mathrm{rel},\mathrm{zm}}
	(X,\lambda^z\tau;Y,0)
	\\
	&=
	\mathcal P_r\mathcal S_\lambda
	C_{2,2}^{\mathrm{rel},\mathrm{zm}}
	(X,\lambda^z\tau;Y,0)
	\\
	&=
	\lambda^{\zeta_4}
	F_{2,2}^{\mathrm{rel},\mathrm{zm}}
	(r,\tau).
	\end{aligned}
	\label{eq:app_F22_dynamic_homogeneity}
\end{equation}
Thus, $F_{2,2}^{\mathrm{rel},\mathrm{zm}}$ is a generalized homogeneous function of $r$ and $\tau$ with spatial degree $\zeta_4$ and the intrinsic relative-dispersion exponent $z=2-\xi$.
Given that the coefficient $D_1$ has dimensions of $[D_1] = [\mathrm L]^{2-\xi}[\mathrm T]^{-1} = [\mathrm L]^z [\mathrm T]^{-1}$, the dimensionless time variable is given by
\begin{equation}
	s
	=
	\frac{D_1\tau}{r^z}
	=
	\frac{D_1\tau}{r^{2-\xi}}.
	\label{eq:app_scaling_variable_s}
\end{equation}
Consequently, Eq.~\eqref{eq:app_F22_dynamic_homogeneity} yields
\begin{equation}
	F_{2,2}^{\mathrm{rel},\mathrm{zm}}(r,\tau)
	=
	A_4r^{\zeta_4}
	\varPhi_{2,2}
	\left(
	\frac{D_1\tau}{r^{2-\xi}}
	\right).
	\label{eq:app_F22_scaling_form}
\end{equation}
Here, the amplitude $A_4$ is determined by the equal-time inertial-range relation $F_{2,2}^{\mathrm{rel},\mathrm{zm}}(r,0) = A_4r^{\zeta_4}$, and the scaling function is normalized such that $\varPhi_{2,2}(0)=1$.

Equation~\eqref{eq:app_F22_scaling_form} is a direct consequence of inertial-range scale covariance rather than an independent ansatz.
It fixes the intrinsic relative-dispersion exponent and collapses the dependence on $r$ and $\tau$ into a single scaling variable.
However, scale covariance alone is insufficient to determine the scaling function $\varPhi_{2,2}$.
Determining $\varPhi_{2,2}$ requires the action of the full active generator $\mathcal M_X$ on the four-particle zero mode $Z_4$.
In particular, obtaining a closed radial equation for $\varPhi_{2,2}$ requires establishing the invariance of the projected isotropic sector under $\mathcal M_X$.

\subsection{Proof of Eq.~\eqref{S-eq:varPhi22_sol_exact}}\label{app:supp_exact_Phi22_solution}
We derive the analytic solution to Eq.~\eqref{eq:supp_Phi22_ODE}.
First, we recast Eq.~\eqref{eq:supp_Phi22_ODE} as
\begin{equation}
	2(d-1)z^2s^2\varPhi_{2,2}''
	-
	\left[
	1
	+
	2(d-1)z(2\zeta_4+d-2z)s
	\right]
	\varPhi_{2,2}'
	+
	2(d-1)\zeta_4(\zeta_4+d-z)
	\varPhi_{2,2}
	=
	0 .
	\label{eq:app_Phi22_rearranged}
\end{equation}
Introducing the parameter $\kappa=2(d-1)z^2$ and the variable $x=1/(\kappa s)$, we define the transformed function $\phi(x)=\varPhi_{2,2}(\frac{1}{\kappa x})$.
Under this change of variables, the derivatives transform as
\begin{equation}
	\frac{\dif x}{\dif s}
	=
	-\kappa x^2,
    \quad
    \varPhi_{2,2}'(s)
	=
	-\kappa x^2
	\frac{\dif\phi}{\dif x},
    \quad
    \varPhi_{2,2}''(s)
	=
	\kappa^2
	\bigg(
	x^4\frac{\dif^2\phi}{\dif x^2}
	+
	2x^3\frac{\dif\phi}{\dif x}
	\bigg).
\end{equation}
Substituting $s=(\kappa x)^{-1}$ into Eq.~\eqref{eq:app_Phi22_rearranged} yields
\begin{equation}
	\begin{aligned}
	x^2\frac{\dif^2\phi}{\dif x^2}
	+
	\bigg(
	x^2
	+
	\frac{2\zeta_4+d}{z}x
	\bigg)
	\frac{\dif\phi}{\dif x}
	+
	\frac{\zeta_4(\zeta_4+d-z)}{z^2}
	\phi
	=
	0 .
	\end{aligned}
	\label{eq:app_phi_x_equation}
\end{equation}

The variable substitution
\begin{equation}
	\phi(x)
	=
	x^{m}g(u),
	\quad
	u=-x,
    \quad
    m=-\zeta_4/z
	\label{eq:app_phi_to_g}
\end{equation}
transforms the derivatives of $\phi=x^m g(u)$ as
\begin{equation}
	\frac{\dif\phi}{\dif x}
	=
	x^{m-1}
	\bigg(
	mg
	-
	x\frac{\dif g}{\dif u}
	\bigg),
    \quad
    \frac{\dif^2\phi}{\dif x^2}
	=
	x^{m-2}
	\bigg(
	m(m-1)g
	-
	2mx\frac{\dif g}{\dif u}
	+
	x^2\frac{\dif^2g}{\dif u^2}
	\bigg).
	\label{eq:app_phi_x_derivative}
\end{equation}
Substituting Eq.~\eqref{eq:app_phi_x_derivative} into Eq.~\eqref{eq:app_phi_x_equation}, we obtain
\begin{equation}
	u\frac{\dif^2g}{\dif u^2}
	+
	\left(
	\frac{d}{z}-u
	\right)
	\frac{\dif g}{\dif u}
	+
	\frac{\zeta_4}{z}g
	=
	0,
	\label{eq:app_Kummer_equation_explicit}
\end{equation}
which is a Kummer equation.
Consequently, for generic $d/z$, the general solution is
\begin{equation}
	\varPhi_{2,2}(s)
	=
	c_1
	x^{-\zeta_4/z}
	{}_1F_1
	\bigg(
	-\frac{\zeta_4}{z};
	\frac{d}{z};
	-x
	\bigg)
	+
	c_2
	x^{-(\zeta_4+d-z)/z}
	{}_1F_1
	\bigg(
	-\frac{\zeta_4+d-z}{z};
	2-\frac{d}{z};
	-x
	\bigg),
	\label{eq:app_Phi22_general_solution}
\end{equation}
with coefficients $c_1$ and $c_2$ to be determined.
For $d>z$, the second branch associated with $c_2$ contributes a singular term proportional to $r^{z-d}$ and must therefore be excluded.
Consequently, we require $c_2=0$, while the normalization condition $\varPhi_{2,2}(0)=1$ determines $c_1 = \Gamma((d+\zeta_4)/z) / \Gamma(d/z)$.
This yields the normalized scaling function
\begin{equation}\label{eq:app_Phi22_exact_solution}
	\varPhi_{2,2}(s)
	=
	\frac{
	\Gamma\bigg(\dfrac{d+\zeta_4}{z}\bigg)
	}{
	\Gamma\bigg(\dfrac{d}{z}\bigg)
	}
	x^{-\zeta_4/z}
	{}_1F_1
	\bigg(
	-\frac{\zeta_4}{z};
	\frac{d}{z};
	-x
	\bigg).
\end{equation}

\bibliographystyle{apsrev4-2.bst}
\bibliography{supp.bib}